\documentclass[journal,comsoc]{IEEEtran}

\usepackage{cite}

\ifCLASSINFOpdf
  \usepackage[pdftex]{graphicx}
\else
  \usepackage[dvips]{graphicx}
\fi

\usepackage[hyphens]{url}

\usepackage{subfigure}
\usepackage{xcolor}
\usepackage{algorithm}
\usepackage{bbm}
\usepackage{framed}
\usepackage{algpseudocode}
\usepackage{ifthen}
\usepackage{ulem}
\usepackage{color,soul}
\usepackage{epsfig}
\usepackage{epstopdf}
\usepackage{times}
\usepackage{amssymb}
\usepackage{amsmath,amsthm}
\usepackage{mathtools}
\usepackage{multicol}
\usepackage{eurosym}
\usepackage{amsthm}
\usepackage{xspace}
\usepackage{pdfpages}
\usepackage{verbatim}
\usepackage{tabularx}
\usepackage{arydshln}
\usepackage{makecell}
\usepackage{multirow}
\newenvironment{sketch}{{\noindent \it Sketch of Proof:~}}

\newtheorem{lemma}{Lemma}

\newtheorem{definition}{Definition}

\newcommand{\version}{paper}
\newcommand{\cond}[2]{\ifthenelse{\equal{\version}{#1}}{#2}{}}
\makeatletter
\renewcommand{\ALG@beginalgorithmic}{\small}
\makeatother

\newcommand{\change}[1]{\textcolor{blue}{#1}}

\newcommand{\vv}[1]{\boldsymbol{#1}}

\renewcommand{\emph}[1]{\textit{#1}}
\begin{document}

\markboth{Submitted to IEEE Transactions on Network and Service Management}%
{Shell \MakeLowercase{\textit{et al.}}: Bare Demo of IEEEtran.cls for IEEE Communications Society Journals}


\title{z-TORCH: An Automated NFV \\Orchestration and Monitoring  Solution}


\author{Vincenzo~Sciancalepore,~\IEEEmembership{Member,~IEEE,}
        Faqir~Zarrar~Yousaf,~\IEEEmembership{Member,~IEEE,}
        and~Xavier~Costa-Perez,~\IEEEmembership{Member,~IEEE}
\thanks{V. Sciancalepore, F. Z. Yousaf and X. Costa-Perez are with NEC Laboratories Europe GmbH, Heidelberg, Germany, E-mails: \{vincenzo.sciancalepore, zarrar.yousaf, xavier.costa\}@neclab.eu.}
}

\maketitle

\begin{abstract}
Autonomous management and orchestration (MANO) of virtualized resources and services, especially in large-scale Network Function Virtualization (NFV) environments, is a big challenge owing to the stringent delay and performance requirements expected of a variety of network services. The Quality-of-Decisions (QoD) of a Management and Orchestration (MANO) system depends on the quality and timeliness of the information received from the underlying monitoring system. The data generated by monitoring systems is a significant contributor to the network and processing load of MANO systems, impacting thus their performance. This raises a unique challenge: \emph{how to jointly optimize the QoD of MANO systems while at the same minimizing their monitoring loads at runtime?} This is the main focus of this paper. 

In this context, we propose a novel automated NFV orchestration solution, namely z-TORCH (zero Touch Orchestration) that jointly optimizes the orchestration and monitoring processes by exploiting machine-learning-based techniques. The objective is to enhance the QoD of MANO systems achieving a near-optimal placement of Virtualized Network Functions (VNFs) at minimum monitoring costs.
\bigskip
\end{abstract}

\begin{IEEEkeywords}
NFV, VNF, Orchestration, MANO, Monitoring, Function placement.
\end{IEEEkeywords}

\thispagestyle{empty}

\IEEEpeerreviewmaketitle

\vspace{-4mm}
\section{Introduction}
\label{sec:introduction}

\IEEEPARstart{N}{etwork} Function Virtualization (NFV) is widely being considered as one of the key enabling technologies for upcoming 5G networks. One of the main motivating factors behind NFV is to provide a technology that will enable the operators and service providers to provide and manage resources and services in an efficient and agile manner with reduced CAPital Expenditure (CAPEX) and OPerational EXpenditure (OPEX), reduced new service roll-out time and increased Return-On-Investment (ROI). 

\begin{figure}[!h]
\centering
 \includegraphics[clip=true, scale=0.5]{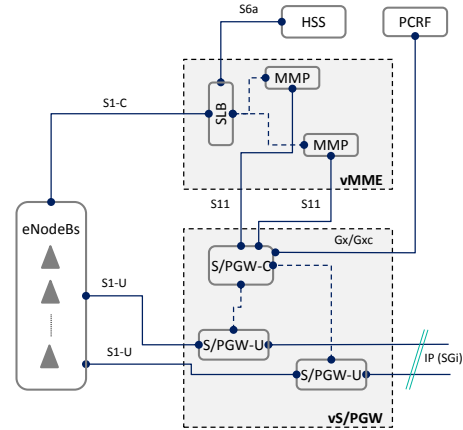}
 \caption{Example of a vEPC VNF with its respective VNF Components.}
 \label{fig:vepc}
 \end{figure}

An NFV system consists of Virtualized Network Functions (VNFs) that are deployed on servers, commonly referred to as compute nodes, located inside the data-center. A Cloud Management System (CMS) is an integral part of such an NFV Infrastructure (NFVI) that is responsible for the Management and Orchestration (MANO) of NFVI resources, such as compute nodes, CPU, network, memory, storage, VNFs etc. For effective MANO decisions, the CMS relies on the presence of a reliable and robust monitoring system that monitors the utilization of the NFVI resources and VNF Key-Performance Indicators (KPIs) and keeps the CMS updated by the regular provisioning of such information. The CMS regularly analyzes the monitored data and derives appropriate Lifecycle Management (LCM) decisions. According to a conservative estimate, up to $25\%$ of enterprise data today is from systems monitoring, with almost $240$ terabytes produced annually~\cite {Kutare2010}. This is likely to grow many folds with the wide deployment of NFV. The challenge thus is \emph{to achieve optimum MANO decisions with reduced monitoring load}. 

\subsection{CMS operational mode}
As part of the MANO operations, the CMS imparts relevant LCM actions on the individual VNFs and its underlying resources in order to ensure its operational and functional integrity. LCM actions may include scaling-in,-out,-up,-down, migration, update or upgrade, delete, ect., of individual VNFs and its respective resources. Providing correct LCM decisions is by itself a challenging problem owing to the variety of VNFs that needs to be managed inside an NFVI. The complexity of a VNF may also vary as advanced VNFs may embody a complete system, for example a virtualized EPC (vEPC) system that is formed of multiple VNF components (VNFC) interlinked over standard and proprietary virtual links. The example of such a complex VNF is illustrated in Fig.~\ref{fig:vepc} \cite{vepcDeployCost}. The MANO complexity of a CMS further increases when it manages Network Services (NS), i.e., designed chains of relevant VNFs, e.g. firewalls, video optimizers, schedulers, virtualized EPCs, etc.

LCM decisions on actions involving resource elements, if not taken with care and deliberations,
may have an inadvertent adverse impact on other resource elements that may be relying on shared services. For example, a migration decision on a VNF belonging to a particular active NS may not only have an adverse impact on the overall QoS of the NS itself but, it may also inadvertently exacerbate the QoS of other VNFs that may be sharing resources with the migrated VNF due to resource contention. Thus, the QoS degradation of one NS may also impact on the QoS of all other NSs relying on the services offered by that particular NS. Therefore, the CMS performs a second iteration of LCM actions to rectify from degraded service situations. This may incur in multiple iterations before the optimal state is achieved. However, multiple iterations of LCM decisions within a short span of time might result in continuous service interruptions thereby impacting the overall QoS/QoE. In other words, the CMS exhibits a poor Quality-of-Decisions (QoD). 

\subsection{The Quality-of-Decisions}
\label{s:qod}
The notion of Quality-of-Decisions (QoD) was pioneered in~\cite{ravaConf} as an indicator of the effectiveness of CMS in terms of imparting MANO decisions. In particular, the QoD is measured in terms of the following mutually dependent criteria: 
\begin{enumerate}
	\item The efficiency of the management action. The resource efficiency is in turn measured in terms of:	
	\begin{itemize}
		\item Whether both the long-term and short-term resource requirements of the managed VNF is fulfilled in the selected compute node;
		\item How non-intrusive a management action has been for other VNFs that are already provisioned in the selected compute node. 
	\end{itemize}
	\item Number of times the management action has to be executed before the most-suitable compute node is determined to migrate or scale the managed VNF.
\end{enumerate}

The QoD of the CMS in turn depends on both the \emph{quality} and \emph{quantity} of the information that it receives from the monitoring system. The quality depends on the variety of KPIs that is reported to the CMS whereas the quantity depends on the frequency of KPI updates that the CMS retrieves. Information provided by a monitoring system may include a variety of KPIs, e.g., percentage-utilization of specific resource units and aggregate resource utilization values of all the VNFs in a physical machine, load experienced by individual VNFs, other QoS parameters, etc. The CMS may then analyze the received data in order to find the state of the NS and take appropriate LCM actions, for e.g. whenever it senses high-utilization events. Moreover, a CMS may manage and orchestrate services that span across multiple data-centers that are geographically apart~\cite{etsiIfa022} and thus rely on receiving monitored data from all the data-centers that are under the CMS administrative domain. However, the problem being that considering the size of an NFVI, where a single NFVI-PoP may host $100$s of $1000$s of compute nodes and, each compute node may host $10$s of $100$s of VNFs and thus, \textcolor{blue}{the CMS ends up managing} $1000$s of VNF instances. The scale of the assets that the CMS requires to monitor further increases in case of multiple data-centers.

Taking into consideration the scale of the resources monitored by the CMS results in a very high load that must be delivered periodically by the monitoring system thereby leading to a high processing load due to data processing and analysis activities. This also causes a processing delay that may result in sluggish reaction to unwanted events. Even with the provisioning of sufficient monitored data, the QoD of the CMS cannot be guaranteed as it depends also on the intelligence of the orchestration algorithm that exploits data from the monitoring system. 

\subsection{Objectives and contributions}
The challenge is thus to jointly optimize both the CMS orchestration process and monitoring process. In this paper, we propose a novel orchestration mechanism, which we refer to as zero-Touch ORCHestration (z-TORCH) method \change{that autonomously enhances the QoD of the CMS orchestration logic at minimum monitoring load during run-time operations. The challenge becomes all the more complex considering the multi-dimensional nature of the cloud infrastructure with a variety of KPIs and resources resulting in a myriad of permutations. Therefore, we address such issue by employing a machine-learning-based method. In particular, we rely on two different techniques: the former is the \emph{unsupervised learning} for processing ``unlabeled'' data about the monitored VNF KPIs so as to efficiently cluster them into accurate VNF profiles, the latter is the \emph{reinforcement learning} to iteratively find a trade-off between solution reliability and complexity (and overhead) of the monitoring system.
} 

The contributions of our paper can be summarized as follows: $i$) we propose an unsupervised binding affinity process in order to profile the VNF KPIs, unveil the correlations between VNF behaviors and group them into VNF affinity groups, $ii$) we analytically study the complexity of our z-TORCH solution and empirically evaluate its convergence properties, $iii$) we devise an adaptive mechanism to dynamically change the number of affinity groups and properly tune the accuracy of the unsupervised binding process, $iv$) we adjust the CMS monitoring frequency based on VNF statistical information by means of the Q-learning theory, $v$) we use a commercial virtualized EPC to configure our VNF profiling for performance evaluation purposes, and $vi$) we show via an exhaustive simulations campaign that z-TORCH exhibits near-optimal performance at low monitoring costs.

The rest of the paper is organized as follows. The next Section~\ref{sec:related_work} gives an overview of the related work. This is followed by Section~\ref{sec:model} providing the detailed description of the system model and the overall z-TORCH architecture. Section~\ref{sect:profiling}, Section~\ref{sect:learning} and Section~\ref{sect:qlearning} show the algorithmic details of our VNF profiling process, VNF placement optimization solution and adjustable monitoring load, respectively. Section~\ref{sec:perf_eval} provides the details of our simulation environment and the performance analysis of the proposed z-TORCH method. We also propose options for the practical deployment of our proposed method in a standard CMS, which is the ETSI NFV MANO system~\cite{etsiMano} in Section~\ref{sec:deployment}. Last, we present a summary of our work and analysis in Section~\ref{sec:conclusion}.

\section{Related Work}
\label{sec:related_work}
The work presented in this paper focuses on the joint optimization of VNF orchestration and monitoring process. \change{There are three main modes---in terms of monitoring process---through which the CMS may receive monitored data: \emph{Periodic Mode} that enables periodic delivery of monitored data, where the period and type of data is specified, \emph{Pull Mode} that provides monitored data only when solicited by the CMS, and \emph{Push Mode} that sends monitored data only when a specific event is triggered, for e.g. CPU burden or when a network load on a VNF exceeds some specific threshold.}
 
While those methods, and combination of them \cite{pushpull2010}, have been exhaustively explored in the literature, they present significant limitations. Periodic reports are identified as the straightforward approach to keep monitoring the resources status but, in case of very large data-centers, it considerably exacerbates the burden and complexity of the monitoring process. Conversely, pull requests option solves the huge overhead issue but it needs a proper design in order to provide the QoE/QoS guarantees and may make the CMS miss out on some critical events. Lastly, the push mode can be tuned so as to recover the system when it is close to alert-states but it may prevent from an optimal allocation/distribution of VNFs within the available compute nodes. Thus, none of these three traditional techniques offer a reliable and optimized solution for large scale NFVI-PoPs and their shortcomings have an adverse impact on the CMS' QoD. Therefore, there is an impelling need to develop an adaptive approach where the monitoring system can adapt according to the events.

In terms of adaptive monitoring systems, there are proposals related to adaptive sampling especially in the domain of wireless networks where energy, processing and bandwidth resources are at premium. Some of them utilize learning techniques like reinforcement learning to make optimum choice of sampling data. Typically such approaches would include clustering, data aggregation and prediction to determine the data sampling frequency. For example \cite{borgne2007} proposes an adaptive model selection (AMS) algorithm that relies on a-priori knowledge of models which is used by the sensor to compare its real measurement with the predicted ones, and only communicate data in case of large variance between the measured and predicted values. This saves on the communication load but it is still computationally expensive as the sensors need to continually sample measurements besides other shortcomings. \cite{Malik2011} optimizes the query method of GWs for collecting periodical data from the monitored objects by employing a statistical technique called principal component analysis on historical traces of sensory data to automatically identify sensors that measured most of the variance observed in the environment. Data from only those sensors would then be collected reducing the transmission cost by up to 50\%. This approach however does not take into account unpredictable environmental evolution yielding inaccurate data. Such a method is not feasible owing to the more frequent unpredictable workload variation on VNFs inside the NFVI. Another proposal is~\cite{Gabriel20172} that employs single rule defining the sampling interval according to the Time of Day, where sampling frequency is high during busy hours periods. It also employs Dual Prediction scheme (DPS) for prediction outside the busy hour based on historical data. This method again cannot be relied on in large scale NFV environment where multiple NS may exist with a different busy hour definition. A more recent work reported in \cite{Gabriel20171} proposes a dynamic sampling rate adaptation scheme based on Reinforcement Learning that is able to tune temperature sensors’ sampling interval on-the-fly, according to environmental conditions and application requirements. The optimization goal is to avoid oversampling and save energy. The method selects from a predefined set of sampling frequencies making it unsuitable for the more dynamic and multi-variable NVF environment. Moreover, adaptive sampling methods usually focus on intelligently varying sampling frequency but ignore the duration of the surveillance epoch. Both these factors are crucial in NFV environment as the CMS is supposed to consider a LCM decision at the end of each surveillance epoch.

In the context of NFV orchestration, a large library of works is present proposing different VNF placement algorithms with different optimization goals. Each proposed solution is unique to its own problem space and use case. We only present some of the more recent works in order to give an overview of the prevailing trends and needs of the industry in this very important problem space.

The authors in \cite{talebVnfp2015} propose VNF placement algorithms with two-fold objective of minimizing path between users and data anchor gateways, and optimizing the sessions' mobility. In \cite{sunVnfp2016} the authors propose a time-efficient heuristic based on affiliation-aware VNF placement for NS deployment. It also proposes an on-line forecast-assisted NS deployment algorithm that predicts the future VNF requirements. For optimizing the VNF placement decisions in response to on-demand workload, \cite{ghaznaviVnfp2015} proposes a solution called Simple Lazy Facility Location (SLFL) that results in the doubling of workload acceptance while incurring similar operational costs compared to first-fit and random placements. \cite{carpioVnfp2017} explores the problem of VNF placement problem in the context of network load balancing in data-centers. It explores the placement of VNFs in smaller clusters of servers in the network thus minimizing the distance-to-the-data-center problem while considering the resources utilization. The authors study the problem of VNF placement with replications to help load balance the network. They design and compare three optimization methods, including Linear Programing (LP) model, Genetic Algorithm (GA) and Random Fit Placement Algorithm (RFPA) for the allocation and replication of VNFs showing significant improvement in load balancing. In the context of enterprise WLAN, \cite{riggioVnfp2015} proposes a VNF placement algorithm for optimizing the functions deployment according to application level constraints. The proposal depends on the presence of hybrid nodes that combine the forwarding capabilities of a programmable switch with the storage/computational capabilities of a server. On similar trends, \cite{frangoVnfp2016} studies the on-demand deployment of VNFs in telco CDNs. 

All of the above cited work tackle the VNF placement problem with a narrow viewpoint of a particular use case with specific requirements. However, there is a need to have a more universal approach to the VNF placement problem in particular and NFV Orchestration in general. Moreover, none of the above proposals take into account the orchestration cost and the monitoring load. The only work that does consider the CMS orchestrator QoD is one of our earlier works in~\cite{ravaConf,ravaMag}, in which we present a Resource Aware VNF Agnostic (RAVA) orchestration method that employs a very different approach of using Pearson Correlation method for optimum placement of VNFs with reduced orchestration cost. This method relies heavily on the frequent provisioning of monitored data from the underlying monitoring system. Moreover, the method does not provide an accurate VNF profile, which is a crucial aspect of the placement decisions. In view of this, we propose z-TORCH that jointly optimizes the NFV orchestration and monitoring process so as to achieve near-optimal placement of VNFs at reduced monitored load while enhancing the CMS' QoD.

\section{System concepts and model}
\label{sec:model}
\begin{figure}[t!]
 \includegraphics[clip,width=\linewidth]{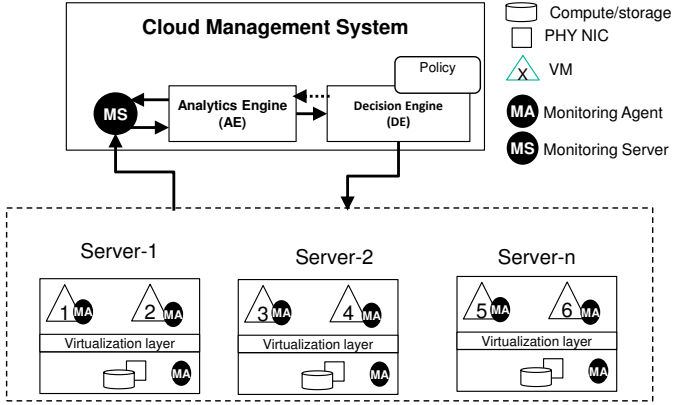}
 \caption{Generic cloud management system.}
 \label{fig:cms}
 \end{figure}
We consider a generic cloud system, as the one depicted in Fig.~\ref{fig:cms}. The infrastructure consists of multiple servers (referred to as compute nodes) and VMs deployed in each server. The server resources (e.g., compute, network, storage, memory) are virtualized and allocated to each VM based on the respective VM requirements. A VM when configured with some network function is referred to as a VNF, or when configured with some application function it is referred to as a Virtualized Application Function (VAF)~\cite{vafMec}. For the sake of clarity, we consider only VNFs in our analysis. A VNF may belong to one or more virtual service instance and the CMS is supposed to manage and orchestrate the infrastructure resources in order to ensure service integrity and to ensure that each VNF is able to fulfil the operational and functional needs of the respective configured application or function. 

To achieve that, the CMS relies on an advanced monitoring system. Given the plethora of available options, we consider in our work a monitoring system called Zabbix~\cite{zabbix}, where the Monitoring Server (MS) is deployed and configured within the CMS, and the MS interacts with one or more Monitoring Clients (MC). The MC instances are distributed within the infrastructure and each MC instance is associated with the entity, for example a VNF and a compute node that needs to be monitored. A MA is an agent for the MS that simply monitors, samples, collects and record the relevant metrics providing them back to the MS. The MS after pre-processing phases passes the data to the Analytics Engine (AE) that processes the monitored data and provides the required analysis output to the Decision Engine (DE). The DE then takes some relevant LCM decisions based on some prescribed policy. Please note that the DE informs the AE with its decision choice and, based on that, the AE is able to derive and provide suitable configuration parameters to the MS for future monitoring rounds. 
Finally, the MC is configured via MS by specifying the relevant metrics and KPIs that the MC is supposed to monitor and record. The MC is also configured with the monitoring granularity (or frequency of monitoring samples).

\subsection{z-TORCH: A dynamic monitoring and deciding process}
Our proposal allows to re-think the classical CMS monitoring and function placement process by introducing a machine-learning-based approach. In particular, we devise a new solution able to $i$) properly select and monitor relevant VNF KPIs, $ii$) evaluating them based on prior (learned) information, $iii$) optimally place them into available compute nodes to keep the system within stable working conditions, $iv$) derive and schedule the next monitoring and decision time instants based on VNFs behaviours information. While this entire framework might appear complex and over-demanding, however it distributes the complexity of a centralized solution over MCs that are dynamically configured. 
\begin{figure}[t!]
 \includegraphics[clip=true,width=0.5\textwidth]{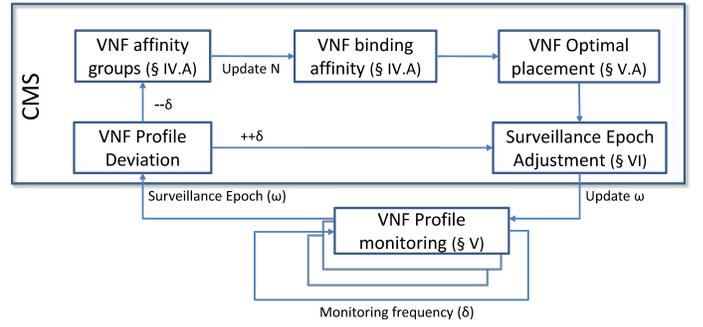}
 \caption{Functional blocks of the z-TORCH solution.}
 \label{fig:dmd}
 \end{figure}

\begin{figure*}[t]
 \centering 
 \includegraphics[clip,width=0.9\textwidth]{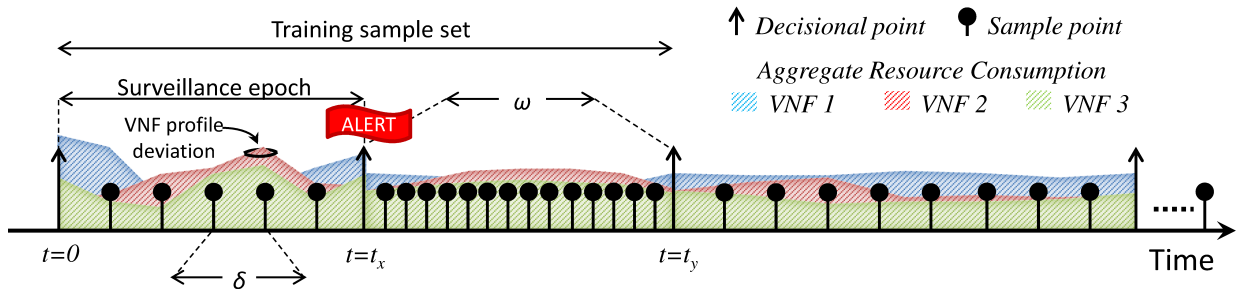}
 \caption{Time evolution of z-TORCH.}
 \label{fig:epoch}
\end{figure*}

Fig~\ref{fig:epoch} shows the overall process. We define a \emph{Surveillance Epoch} ($\omega$)\footnote{\change {To avoid notation clutter, we have removed index ($\tau$) from $\omega^{(\tau)}$. However, a formal definition of $\omega^{(\tau)}$ is provided in Section~\ref{sect:qlearning}.}} as the time window within which we monitor the VNFs' KPIs. Monitoring operations are performed at different \emph{Sample points}, spaced by $\delta$. Surveillance epochs last $\omega$ and are delimited by \emph{decisional points} defined as points in time where our solution takes LCM decisions. In particular, LCM decisions might comprise: $i$) changing the frequency of monitoring information ($\delta$), $ii$) changing the length of the surveillance epoch ($\omega$), $iii$) optimal placement of VNFs based on \emph{unsupervised binding affinity} calculation.
When a critical resource shortage is detected, an alert message is captured by a sample point so that the next decisional point may be expedited to handle unexpected network changes.

\subsection{General solution overview}
\label{sect:system_beh}
Our novel concept of self-monitoring and proactive function placement relies on the concept of an adjustable monitoring frequency based on the machine-learning paradigm. Fig.~\ref{fig:dmd} provides an example of the general process indicating the relevant sections where the respective process is described. At the beginning, the decisional point requires the CMS to generally place VNFs onto available compute nodes. This initial operation might be performed without any a-priori information, namely \emph{blind-placement}, or with some previous information gathered during a training phase. After placing VNFs, the CMS decides the frequency of sample points, i.e., the frequency of monitoring requests each server feeds back to the CMS. This directly affects the overall monitoring overhead that might be unaffordable when facing with thousands of VNFs~\cite{Kutare2010}. In addition, the CMS dynamically decides the length of the surveillance epoch based on a reward function obtained, as explained in Section~\ref{sect:qlearning}. 

The KPIs of any single VNF are identified, based on VNF descriptors available beforehand~\cite{etsiMano}, and are processed. This helps to provide an accurate \emph{profile} of each VNF running in our system, as described in Section~\ref{sect:profiling}. While the number of KPIs may consistently grow, our solution proposes an unsupervised binding affinity calculation to properly find out the correlation among them for any specified VNF. For the sake of simplicity, we consider the generic KPIs for any VNF, such as \emph{Network Load}, \emph{Computational Burden} and \emph{Storage Utilization}~\footnote{While the number of KPIs might be consistent, our solution still provides reasonable results when compared against state-of-the-art solutions, as shown in the next sections.}. A clear example is represented by a firewall VNF. It might be characterized by a high network demand and high storage utilization whereas it might exhibit low computational burden. Affinity values, which indicate the correlation among different VNF profiles, are gathered by the CMS, which can optimally place the VNFs into compute nodes while keeping the overall load of any single compute node in balance. When the functions placement occurs (at $t=0$ in Fig.~\ref{fig:epoch}), a default monitoring frequency $\delta$ and surveillance epoch $\omega$ are fixed. At the next decisional point ($\omega$), the CMS detects any VNF differing from the prior profile information, namely \emph{VNF profile deviation}. This automatically forces the CMS to increase the monitoring frequency in order to anticipate any unexpected critical event, such as compute node resources outage. At the next decisional point, if no other VNF profile deviation has occurred, the monitoring frequency is reduced and the reward function is increased (as explained in the Section~\ref{sect:qlearning}), which, in turn, enlarges the surveillance epoch $\omega$. Conversely, if additional VNF profile deviations have occurred, a new VNF profiling is performed based on the \emph{Training Sample Set}. In this case, the monitoring frequency is restored to a default value and the Surveillance epoch length is reduced (as the reward function is decreased).

\section{VNF profiling process}
\label{sect:profiling}

VNF characteristics can be efficiently analyzed with the aim of properly profiling the resource utilization. In particular, we rely on machine-learning-based techniques to learn from general behaviours so as to be proactive in case of compute node resources shortages.

\begin{figure}
 \includegraphics[clip,width=\linewidth]{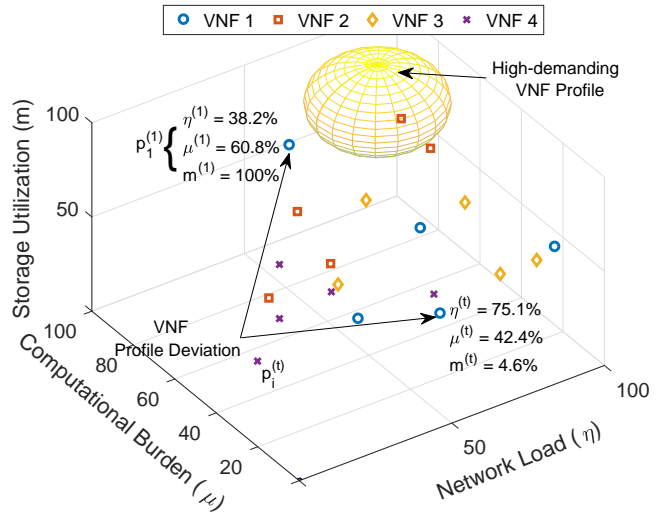}
 \caption{Characterization of VNF profiles in a $3$-dimensional space.}
 \label{fig:3d_pic}
 \end{figure}

\change{We can define the vector $\vv{p}_i^{(t)} = \{m_i^{(t)},\mu_i^{(t)},\eta_i^{(t)}\}$, where $p_i^{(t)}$ is the set of monitored information for each VNF $i$ running in our system at time $t$ whereas $m,\mu,\eta$ are the utilization of storage, CPU and network resources, respectively.} This vector can be depicted as a single-point in a $3$-dimensional space $\mathbb{S}$ within a snapshot of time $t$. In Fig.~\ref{fig:3d_pic}, we show different VNF profiles at consecutive time-snapshots to give a clear idea of our model. The plotting space can be partitioned to identify zones with specific profile properties. For instance, we have highlighted with a yellow sphere the zone wherein VNFs are marked as high-demanding: \change{the main idea is to leverage on such profile partitioning in order to proactively place VNFs into compute nodes while keeping the system stable, i.e., when it does not require further VNF migrations that, in turn, results in high Quality-of-Decisions (as explained in Section~\ref{s:qod})}.

\subsection{Unsupervised binding affinity}
\label{sect:binding}
After defining our modelling space $\mathbb{S}$, we need to characterize different areas based on some peculiarities of all gathered VNF profiles. Without loss of generality, we truncate the index $(t)$ from $p_i^{(t)}$ when not needed. Given a number of affinity profile groups $N=|\mathcal{N}|$, our problem can be formalized as the following: 
\emph{Finding non-overlapping affinity profile groups $n\in\mathcal{N}$ such that $i$) the union set of those groups is equal to the VNF profile space $\mathbb{S}$, $ii$) each affinity group contains at least one element $\vv{p_i}$, $iii$) all VNF profiles $p_i$ must be placed in one VNF profile group.} 

Each VNF profile group $n\in\mathcal{N}$ is characterized by a center of gravity $\vv{c_n} = \{c_{n,(1)}, c_{n,(2)}, \cdots, c_{n,(z)}\}$, where $z\in\mathcal{Z}$ is the spatial dimension ($Z=|\mathcal{Z}|=3$ in our example). The center of gravity of group $n$ is obtained as the spatial point with the least Euclidean distance from all other VNF profile values $p_i$ associated to that group $n$. Mathematically, we can formulate the optimization problem as the following

\vspace{2mm}\noindent \textbf{Problem}~\texttt{VNF-Affinity}:
\begin{equation*}
\label{pr:affinity}
\begin{array}{ll}
\text{minimize}  & \sum\limits_n^{|\mathcal{N}|}\sum\limits_i^{|\mathcal{I}|} x_{i,n}(||\vv{c_n} - \vv{p_i}||_2)\\
\text{subject to } & \sum\limits_n x_{i,n} = 1, \quad\forall i\in\mathcal{I};\\
				   & \sum\limits_i x_{i,n} \geq 1, \quad\forall n\in\mathcal{N};\\
				   & \vv{c_n}\in \mathbb{R}^{|\mathcal{Z}|};\\
				   & x_{i,n}\in \{0,1\},
\end{array}
\end{equation*}
where the outputs are $\vv{c_n}$ defining the spatial coordinates (KPIs) of each center of gravity, and the binary values $x_{i,n}$ indicating whether VNF $i$ is grouped into affinity group $n$, whereas $||\cdot||_2$ is the Euclidean distance between the center of gravity $\vv{c_n}$ for affinity group $n$ and each VNF profile $p_i$. Specifically, the Euclidean distance depends on the number of KPIs (or spatial dimensions $Z$) considered, i.e., $||\vv{c_n} - \vv{p_i}||_2 = \sqrt{\sum\limits_z^{Z} \left( c_{n,(z)}-p_{i,(z)}\right)^2}$. In our example, it holds that $p_{i,(1)} = m_i, p_{i,(2)} = \mu_i, p_{i,(3)} = \eta_i$.
 
In the following paragraphs, we perform the complexity analysis and explain how our heuristic e$k$m works. Then, we describe the process of calculating the density of the affinity groups $N$ based on the current system status. Please note that the number of affinity groups $N$ is decided beforehand and provided to our heuristic. This shall allow the CMS to automatically cope with unexpected system changes and quickly react to keep the system stable.
%

\vspace{1em}\noindent{\bf Complexity analysis.} While the number of affinity groups is given, Problem~\texttt{VNF-Affinity} might be still untractable and it might not be solved in an affordable time. This is stated in the following theorem.
\begin{lemma}
\label{Theo:np-complete}
Given a number of affinity profile groups $N\geq 2$, multiple VNF $i\geq N$ and multiple KPIs $Z\geq 2$, Problem~\texttt{VNF-Affinity} is NP-Hard.
\end{lemma}

\begin{sketch}
We consider $Z=2$ KPIs and $N=2$ affinity profile groups. It is clear that the problem falls in NP. We can apply a polynomial reduction to the well-know graph $k$-coloring problem (\!\!\cite{GJ79}). 
In particular, we are given an instance of the graph $G(V,E)$ wherein vertices are VNF profiles $V=\{1,2,i,\ldots,I\}$ and edges are placed between two points with the largest distance. Therefore, we can formulate the following problem: \change{given $k$ available colors, is there any graph coloring solution that assigns different colors to vertices connected with the same edge?} Assuming that this problem is NP-Complete and considering the color of each vertex as an affinity group, we can state that this problem is reducible to Problem~\texttt{VNF-Affinity} in a polynomial time and thus, Problem~\texttt{VNF-Affinity} is NP-Hard.
When considering multiple affinity profile groups $N\geq 2$, it is hard to place the edges in the $k$-coloring graph~\cite{voronoi_k_cluster}, making the problem even harder. When considering multiple affinity groups and multiple KPIs $Z\geq 2$, it is even more difficult to find a solution to Problem~\texttt{VNF-Affinity}, which proves that the NP-Hardness is rather strong.
\qed
\end{sketch}

\noindent{\bf Enhanced K-means heuristic.}
When dealing with NP-Hard problem, a fast and reasonable heuristic is needed to boil down the complexity of a greedy-search solution. There is a large library of work that address this problem, but we focus on a $k$-means heuristic~\cite{k_means} solving the problem within $O(I\cdot N\cdot Z\cdot c)$ time-complexity~\cite{k_means_slow}, where $c$ is the number of rounds to converge as explained next.

\begin{algorithm}[t]
\centering
\vspace*{-1pt}
\begin{framed}

\begin{enumerate}
\small
\vspace*{-5pt}
\item Initialise $t=0$ and $x_{i,n}=0, \forall i\in\mathcal{I},n\in\mathcal{N}$.
\item Initialise set $\vv{c_n}\!\!^{(t)}$ by using the VNF profiles classification.
\item Update $\Delta^{(t)} =\frac{100}{2^{|\mathcal{I}|}}\cdot t^{\sqrt{I}}$.
\item Apply a grid of points $\vv{w_s}\in\mathcal{S}$ on the VNF profile space $\mathbb{S}$ such that $||\vv{w_\xi}-\vv{w_\zeta}||_2=\Delta^{(t)}, \forall \xi\neq\zeta,(\xi,\zeta)\in\mathcal{S}$.
\item Set $x_{i,n}^{(t)}=1 : n = \text{arg}\min\limits_{n}\, ||\vv{c_n}^{(t)}-\vv{p_i}||_2, \forall i\in\mathcal{I}$.
\item Calculate the center of gravity set \\ $\vv{c_n}^{(t+1)} = \frac{1}{\sum_i x_{i,n}^{(t)}} \sum\limits_{i\in\mathcal{I}} \left( \vv{p_i}\cdot x_{i,n}^{(t)} \right),\quad \forall n\in\mathcal{N}$.
\item Update the center of gravity set based on the nearest grid point $\vv{c_n}^{(t+1)}\!\!=\vv{w_s} : s=\text{arg}\min\limits_{s\in\mathcal{S}} ||\vv{w_s}-\vv{c_n}||_2, \forall n\!\in\!\mathcal{N}$.
\item If $\vv{c_n}\!\!^{(t+1)}\!\neq\!\vv{c_n}\!\!^{(t)}$ then increase $t\!=\!t+1$ and jump to ($3$).
\end{enumerate}
\vspace*{-10pt}
\end{framed}
\caption{Enhanced $k$-means (e$k$m)}
\label{algo:ekm}
\vspace*{-8.5pt}
\end{algorithm}

We rely on the classical definition of $k$-means algorithm and improve it to handle the complexity of our VNF affinity modelling~\cite{cluster-based-ekm}. The main idea behind the well-known algorithm is to devise an iterative-algorithm able to randomly select the centres of gravity $\vv{c_n}$ (regardless of the number of spatial dimensions $Z$) and partition the whole space based on the nearest distance rule from each of those centres $\vv{c_n}$. Iteratively, the algorithm recomputes the new centres of gravity based on the current group member properties $\vv{p_i}$ and apply again the partitioning process until the centres of gravity do not change their positions. As proved by~\cite{k_means_slow}, in the worst-case the algorithm might take up to $2^{\Omega(\sqrt{I})}$ steps to converge. We enhance the performance of such an algorithm by applying a regular grid on the affinity space $\mathbb{S}$, namely enhanced $k$-means (e$k$m) algorithm. Points $w_s\in\mathcal{S}$ of the grid are equally spaced from each other. We then constrain the centres of gravity of each VNF affinity group to reside on some specific spatial points. The granularity of such grid span, i.e., the distance $\Delta$, drives the speed of our algorithm and may be dynamically adjusted to speed up the process while keeping the accuracy of the found solution. This is performed by a step-function $\Delta^{(t)} =\frac{100}{2^{|\mathcal{I}|}}\cdot t^{\sqrt{I}}$: the more the steps to converge, the higher the slope of the step-function. Practically, we design a step-function which grows slowly during the first steps (depending on the number of VNF profile points $I$, i.e., the more the points, the slower the growth) and then, it exponentially grows as the number of steps becomes consistent. This helps the system to find a very accurate solution in the first steps, while forcing the algorithm to quickly converge if the number of steps is high. 

The algorithm pseudo-code is provided in Alg.~\ref{algo:ekm}. To avoid the effects of randomness and to make our solution more efficient, we initialize the set of centres of gravity $\vv{c_n}$ (line $2$) based on a VNF profile classification. Interestingly, this classification can be performed by means of external information providing VNF profile templates (in terms of expected KPIs), given a number of VNF affinity groups $N$. For example, when $N=2$ VNF affinity groups are defined, VNF profile templates may influence the initial choice by placing the centres of gravity at $\vv{c_{n=1}} = \{75\%,75\%,75\%\}$ for the high-demanding VNF profiles and at $\vv{c_{n=2}} = \{25\%,25\%,25\%\}$ for the low-demanding VNF profiles. Clearly, such training data may be automatically updated by the infrastructure provider through a monitoring process.

\vspace{1em}\noindent{\bf VNF affinity groups density.}
While e$k$m algorithm can solve and provide a VNF affinity grouping solution within an affordable time, the key-aspect is the number of VNF affinity groups to build. We leverage on the feedback-loop paradigm to design a \emph{controller} in charge of monitoring the system status and triggering a different number of VNF affinity groups when some events occur. The rationale behind is that the affinity grouping procedure may fail and we need to promptly update the number of groups to handle unexpected VNF profile behaviors. Therefore, with some abuse of notation we define the concept of \emph{VNF profile deviation}, as introduced in Section~\ref{sect:system_beh}, as follows:
\begin{definition}
A VNF profile deviation is an event occurring when a VNF profile $p_i^{(t)}$ changes its KPIs from time $t$ to $t+1$ falling into a new VNF affinity group $n\in\mathcal{N}$, i.e., $x_{i,n}^{(t)}\neq x_{i,n}^{(t+1)}$.
\end{definition}

VNF profile deviations give an indication about the accuracy of our affinity grouping process: if the grouping process failed to capture the variance of its members ($\vv{p_i}$), we need to re-run the grouping process assessing the new VNF profile features. This may highly impact on the VNF function placement (as will be discussed in Section~\ref{sect:vnf_placement}) and may result in a service disruption because of a compute node resources outage.

\begin{figure}
 \includegraphics[clip,width=\linewidth]{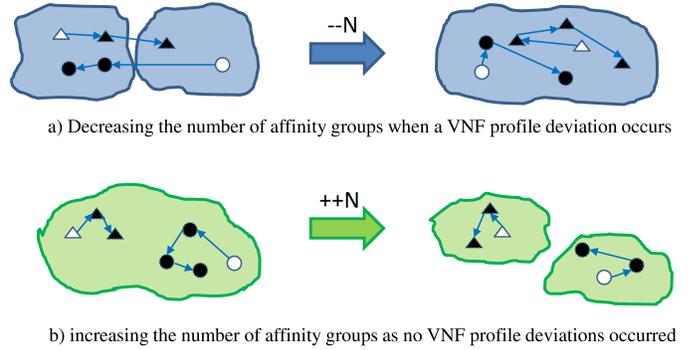}
 \caption{Changing the number of VNF affinity groups based upon VNF profile deviation occurrences.}
 \label{fig:clusters}
 \end{figure}
Fig.~\ref{fig:clusters} shows an example for a $2$-dimensional space, i.e., considering only $2$ KPIs for any VNF profile $\vv{p_i}$. In particular, we show a VNF profile at different times (with solid filled shapes). Please note that those values are snapshots captured at different \emph{sample points}, as explained in Section~\ref{sect:system_beh}. When a VNF profile deviation occurs, an alert message is triggered and more sample points are required (in the next surveillance epoch) to take over the compute node control if some VNF profile exceeds the maximum capacity. At the next decision point, a new VNF affinity binding process is executed and the number of VNF affinity groups $N$ is reduced. This will likely avoid further VNF profile deviations and perform an optimal VNF placement. Conversely, if the VNF profile behavior is predictable and does not exhibit significant changes, after $2$ surveillance epochs the system automatically increases the number of VNF affinity groups to be more accurate in the VNF profiling process. We initially assume the lowest possible number of VNF affinity groups set to $2$. 

\section{VNF placement based on gathered information}
\label{sect:learning}
The number of sample points gathered for the VNF profiling process could significantly affect the VNF placement and, in turn, the overall network performance. Ideally, an infinite number of monitoring samples unveils the correct behavior of such VNF making accurate the VNF profiling. However, this might exacerbate the complexity and the overhead of control messages when applied to a plethora of VNF instances. In our proposal, we trade off the number of monitoring samples against the accuracy of VNF profiling process, which may lead to a huge number of LCM operations, such as migrations or scaling up/down.

Let us consider the realization of a point process $\gamma_{i,(z)} = \sum_{t=0}^T \delta_t p_{i,(z)}(t)$ as the evolution of VNF $i$ and KPI $(z)$, where $\delta_t$ is the Dirac measure for sample $t$. 
\change{
\begin{lemma}
\label{lemma:ergodic}
Given that the VNF profile evolution process is ergodic and stationary, VNF statistical properties can be obtained from any realization of the same process over time or from multiple process instances evaluated at the same time.
\end{lemma}
\begin{sketch}
The proof is rather straightforward. For reasonable short time-lengths of the surveillance epoch, we can consider the VNF profile evolution process as ergodic and stationary, as shown in Section~\ref{s:openepc}. This directly implies that $\bar{\gamma}_{i,(z)} = \frac{1}{K} \sum_{k=0}^K X[k] = \frac{1}{T}\sum_{t=0}^T p_{i,(z)}(t)$, where $k$ are multiple instances of the same process whereas $t$ are different times. This proves the lemma.
\qed
\end{sketch}
}

This lemma helps to significantly reduce the number of decisional points, wherein our VNF affinity binding is executed. In particular, we can collect several profile values of the same VNF (experienced at different sample times) or different VNF instances of the same type to properly characterize a specific VNF profile. Therefore, we use all samples collected within $2$ surveillance epochs in case of an alert message triggered.

\subsection{VNF optimal placement}
\label{sect:vnf_placement}

Once the VNF affinity binding has been successfully completed, the CMS will automatically place VNFs into available compute nodes based on their profile values and their affinity group associations. This being one of main findings of our paper: the objective of our solution is to find an optimal placement that $i$) takes into account the statistical variance of the VNF profile values $\vv{p_i}$, $ii$) places the VNF in order to avoid further LCM operations, such as migrations, $iii$) equally balances compute nodes load to keep the system stable and to reduce the number of monitoring messages (sample points), i.e., to limit the overhead of the monitoring procedure. 

We first apply the VNF placement process to VNF affinity group instances, i.e., considering the center of gravity of each VNF affinity group as a single VNF profile instance. We can formulate the following integer linear programming (ILP) problem

\vspace{2mm}\noindent \textbf{Problem}~\texttt{Proactive-VNF-Placement}:
\begin{equation*}
\label{pr:mt_optimizer}
\begin{array}{ll}
\text{maximize}  & \sum\limits_{l\in\mathcal{L}} \log \sum\limits_{n\in\mathcal{N}} \left( ||\vv{c_n}|| y_{l,n}\right)\\
\text{subject to } & \sum\limits_{n\in\mathcal{N}} \vv{c_n} y_{l,n} \leq \vv{P_l}, \quad\forall l\in\mathcal{L};\\
   				   & \sum\limits_{l\in\mathcal{L}} y_{l,n} \leq 1, \quad\forall n\in\mathcal{N};\\
   				   & y_{l,n}\in \{0,1\},
\end{array}
\end{equation*}
where $||\cdot||$ is the L-$1$ Norm of a vector, $l\in\mathcal{L}$ is an available compute node in our system, $\vv{P_l}=\{P_{l,(z)}\}$ is the set of maximum resource availability for compute node $l$ in terms of KPI $(z)$ whereas $y_{l,n}$ is the binary value indicating whether the VNF class $n$ is placed into compute node $l$. The $\log$ operator is needed to provide fairness between different compute node loads. While Problem~\texttt{Proactive-VNF-Placement} is proved to be NP-Hard\,\footnote{Due to the pages limitation, we skip the formal proof as the problem can be reduced in a polynomial time into a bin packing problem, known to be NP-Hard. However, we refer the reader to~\cite{Kellerer1999} for further details.}, the solution can still be found within an affordable time as the number of variables, i.e., the number of VNF affinity groups $N$, is very limited. In our simulation campaign, we adopt a commercial tool, namely IBM ILOG CPLEX~\cite{ibm_opl}, to solve the optimization problem. 

The solution optimality of Problem \texttt{Proactive-VNF-Placement} can be guaranteed if each VNF profile accurately follows the center of gravity of its assigned affinity group. In other words, the solution optimally works if the bias (variance) from the mean value of the affinity group is very low. Conversely, as soon as the VNF profile values move away from the average properties of its group the scheduling solution might fail leading to unstable system states and service disruptions. Therefore, we devise a VNF scheduling algorithm taking into account the general scheduling information of the VNF affinity groups but applying the current KPIs information of each VNF to correctly balance the compute nodes load.

The pseudo-code of our algorithm, namely Affinity-aided VNF Scheduling (AaVS), is listed in Alg.~\ref{algo:aavs}. Our idea is to rely on the First Fit Decreasing (FFD) algorithm~\cite{FFD_bin_packing}, suggested for bin backing problems. In particular, we calculate (Line $1$) the VNF profile variance as $v_i=\max\limits_{t\in\omega} \left(||\vv{p_i}^{(t)}-\vv{c_n}||_2\right)$ along the last (at least $2$) epochs $\omega$. This is further supported by Lemma~\ref{lemma:ergodic}. Based on variance, each VNF profile value is sorted within its affinity group (Line $4$), leaving at the first position the VNF profile which has experienced much variations (and might be considered as unstable). The rationale behind is that we need to first place the VNF profile which might cause (in the worst case) unexpected compute node resource outages. Iteratively, we try to schedule the other VNF profiles based on Problem~\texttt{Proactive-VNF-Placement} (Line $5$), i.e., based on $y_{l,n}$. Upon all VNFs have been scheduled into compute nodes $l$, our algorithm ends.

\begin{algorithm}[t]
\centering
\vspace*{-1pt}
\begin{framed}
\begin{enumerate}
\small
\vspace*{-5pt}
\item Initialise $v_i=\max\limits_{t\in\omega} (||\vv{p_i}^{(t)}-\vv{c_n}||_2$ : $x_{i,n}=1, \forall i\in\mathcal{I}$.
\item Initialise set $\mathcal{H}_l = \emptyset, \forall l\in\mathcal{L}$, $\mathcal{B}_n = \emptyset, \forall n\in\mathcal{N}$, $\mathcal{F}=\emptyset$ and $l=0$.
\item Place $i\rightarrow\mathcal{B}_n, \forall n\in\mathcal{N}$ if $x_{i,n} = 1$.
\item Sort $\mathcal{B}_n,\forall n\in\mathcal{N}$ according to $v_i$ in a decreasing order.
\item For every $n$, take the first $i$ from $\mathcal{B}_n$ and Place $i\rightarrow\mathcal{F}$ if $y_{l,n}=1$. If $i$ does not fit, Take the next $i$ in $\mathcal{B}_n$.
\item Remove all $i$ placed in $\mathcal{F}$ from $\mathcal{B}_n$. Update $\mathcal{H}_l\leftarrow\mathcal{F}$.
\item If there is any $n:\mathcal{B}_n\neq\emptyset$ then Increase $l=l+1$, Update $\mathcal{F}=\emptyset$ and go to (5).
\end{enumerate}
\vspace*{-10pt}
\end{framed}
\caption{Affinity-aided VNF Scheduling (AaVS)}
\label{algo:aavs}
\vspace*{-8.5pt}
\end{algorithm}

Assuming that the compute nodes deployment is over-provisioning, Problem~\texttt{Proactive-VNF-Placement} can reasonably purse at balancing the load of compute nodes and keep them in a stable state without dangerously approaching the saturation point. Nonetheless, to avoid unexpected compute node resources saturation, $\vv{P_l}$ in Problem~\texttt{Proactive-VNF-Placement} can be properly chosen by the infrastructure provider. When AaVS is applied, the fairness among different compute nodes can significantly degrade because of unpredictable VNF profile spikes. Therefore, we design a controller in charge of promptly changing the number of VNF affinity groups (and re-grouping VNFs profiles) when VNF profiles significantly differ from the VNF affinity group properties, as explained in Section~\ref{sect:binding} and empirically shown in Section~\ref{sec:perf_eval}. However, the entire process could be affected by the length of the surveillance epoch $\omega$, which is dynamically adjusted, as explained in the next section.

\section{Monitoring overhead adjustment}
\label{sect:qlearning}

The decisional points play a key-role because: $i$) at those times the system might re-build the affinity groups and improve the accuracy of the VNF placement that, in turns, translates into a better Quality-of-Decisions (QoD) and less LCM operations in the near future due to a stable system conditions, $ii$) complexity and overhead of the DMD are strictly related to the frequency of the decisional points, i.e., surveillance epoch length $\omega$. 
An optimal trade-off must be found based on the current system conditions as well as previous observations. We design an adaptive scheme to keep track of previous alert triggers while increasing the surveillance epoch when the stability of the system can be preserved for a longer time period.

Our scheme is based on the well-known Q-Learning approach~\cite{Watkins92q-learning}. The main idea behind is to learn from previous actions and obtained rewards in order to take the optimal decision in the future while pursuing the reward maximization. Without loss of generality, we define the index of surveillance epoch as well as the decisional point at the end of a surveillance epoch by $\tau\in\mathcal{T}$. Let us define the state space $\pi\in\Pi$ as the number of VNF profile deviations $j$ experienced at the previous decisional point, i.e., $\pi^{\tau} = j_{(\tau-1)}$. At every decisional point $\tau$, our system may take different actions $a^{\tau}$ on how much to increase (decrease) the next surveillance epoch $\omega^{(\tau+1)}$, i.e., $a^{\tau} = \{+k\cdot o\}$, where $o$ is defined as the least step size. After taking an action $a^{\tau}$, the system will 
be rewarded based on a reward function $R(\pi^{\tau},a^{\tau}) = \frac{\omega^{(\tau)}}{j_{\tau}^\beta}$, where $\omega^{(\tau)}$ is the length of the surveillance epoch between two decisional points $\tau-1$ and $\tau$.
%
The objective is to maximize the surveillance epoch $\omega$ while keeping low (or zero) the number of VNF profile deviations occurred in the last surveillance epoch, which might compromise the stability of our system. $\beta\leq 1$ is a tunable parameter that can be adjusted by the infrastructure provider to have a slower (faster) changing of the surveillance epoch at expense of less (more) scheduling optimality.

Our solution builds a Q-table collecting the reward coming from each possible pair $(\pi,a)$ based on the following equation
\begin{equation}
Q(\pi,a)\! =\! (1-\alpha)Q(\pi,a)\! +\! \alpha\left[R(\pi^{\tau},a^{\tau},\pi^{\tau+1})\! +\! \psi q_{\text{max}}\right],
\end{equation}
where $q_{\text{max}} = \max\limits_{a^{\tau+1}} Q(\pi^{\tau+1},a^{\tau+1})$, and $R(\pi^{\tau},a^{\tau},\pi^{\tau+1})$ is the reward obtained from action $a^{\tau}$ leading to state $\pi^{\tau+1}$.
$\alpha$ and $\psi$ are the learning rate and the discount rate, correspondingly. The former balances the stored information (in the Q-table) against the current observed ones. It is usually set differently per state and evolving over time, i.e., $\alpha^{\tau}_{\pi,a} = \frac{0.5}{i(\pi,a)}$, where $i(\pi,a)$ is the number of times we have explored state $\pi$ by time $\tau$. The latter gives a less weight to old information, which could become incorrect. This is useful when the stationary and ergodic assumption on the VNF statistical properties could not be taken for very long periods (please refer to Section~\ref{sect:learning}). This is commonly fixed to $0.9$ (\!\cite{Watkins92q-learning}). When a new action must be taken, our system may select it randomly (with probability $\phi\leq 1$) among available actions $a\in\mathcal{A}$ or it can select the one maximizing the reward (with probability $1-\phi$) based on the information stored in the Q-table, i.e., $a = \text{arg}\max\limits_{a^\tau} Q(\pi,a)$.

\section{Performance Evaluation}
\label{sec:perf_eval}
We conduct an exhaustive simulation campaign by means of a mathematical tool, such as MATLAB. All building blocks of our solution are implemented and executed using several random seeds to keep the confidence degree of our results below $0.1\%$. To validate our results, we evaluate a realistic use case using virtual functions deployed in our testbed. This provides a set of reference points for our VNF profiles creation process.

\subsection{Evaluation case: OpenEPC}
\label{s:openepc}

\begin{figure}[t]
 \includegraphics[clip=true,width=0.5\textwidth]{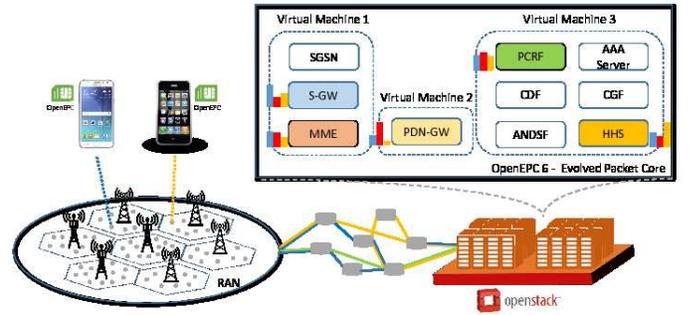}
 \caption{Real Evaluation Case: OpenEPC mobile core.}
 \label{fig:openEPC}
 \end{figure}

We implement a real network deployment with $2$ NEC eNBs~\cite{nec-smallcell} and a virtualized core domain using a commercial software, OpenEPC~\cite{openepc}. Our testbed deployment is shown in Fig.~\ref{fig:openEPC}. Mobile devices provided with a customized SIM-card are connected to the mobile core domain, running on OpenStack.   
Different KPIs for any specific VNF, such as MME, S-GW and P-GW are collected by means of Ceilometer, a telemetry software provided with OpenStack. The evaluation time window is set to $2$ hours and two different user profiles are considered: \emph{high-demanding} in case of data traffic upload and download, \emph{low-demanding} in case of high-mobility (several hand-overs) but no data traffic (only control signal).

\change{Overall results are summarized in Table~\ref{tab:kpis_vnf}. We have classified only the most significant KPIs (in percentage), based on the total capacity of compute nodes. Interestingly, they suggest a specific set of requirements that are considered and exploited throughout our performance evaluation section, ranging from low demanding requirements, e.g., HSS for \emph{low} configuration, up to high-demanding requirements, e.g., PDN-GW for \emph{high} configuration.}

\begin{table}[h!]
\vspace{-3mm}
\caption{Virtualized Network Functions KPIs (OpenEPC 6)}
\label{tab:kpis_vnf}
\vspace{-2mm}
\scriptsize
\centering
\begin{tabular}{| c || c : c|c : c|c : c|}
\hline
 & \multicolumn{2}{c|}{\textbf{CPU ($\mu$) [$\%$]}} & \multicolumn{2}{c|}{\textbf{Mem} (m) [$\%$]} & \multicolumn{2}{c|}{\textbf{Net ($\eta$) [$\%$]}}\\
\textbf{VNFs} & Low & High & Low & High & Low & High\\
\hline
MME & $17.7$ & $2.9$ & $15.9$ & $3.8$ & $5.8$ & $1.9$\\
S-GW & $0.7$ & $79.1$ & $0.3$ & $3.3$ & $0.14$ & $91.2$\\
HSS & $0.9$ & $2.9$ & $1.1$ & $4.5$ & $0.7$ & $1.3$\\
PCRF & $1.2$ & $1.9$ & $0.6$ & $3.9$ & $0.5$ & $0.9$\\
PDN-GW & $1.7$ & $53.1$ & $2.1$ & $37.2$ & $0.8$ & $92$\\
\hline
\end{tabular}
\end{table}

\begin{figure*}[t!]
 \centering
 \subfigure[VNF Affinity Groups (N) varying $I$]
 {
	\label{fig:multiI_N}        
    \centering
	\includegraphics[width=0.3\textwidth]{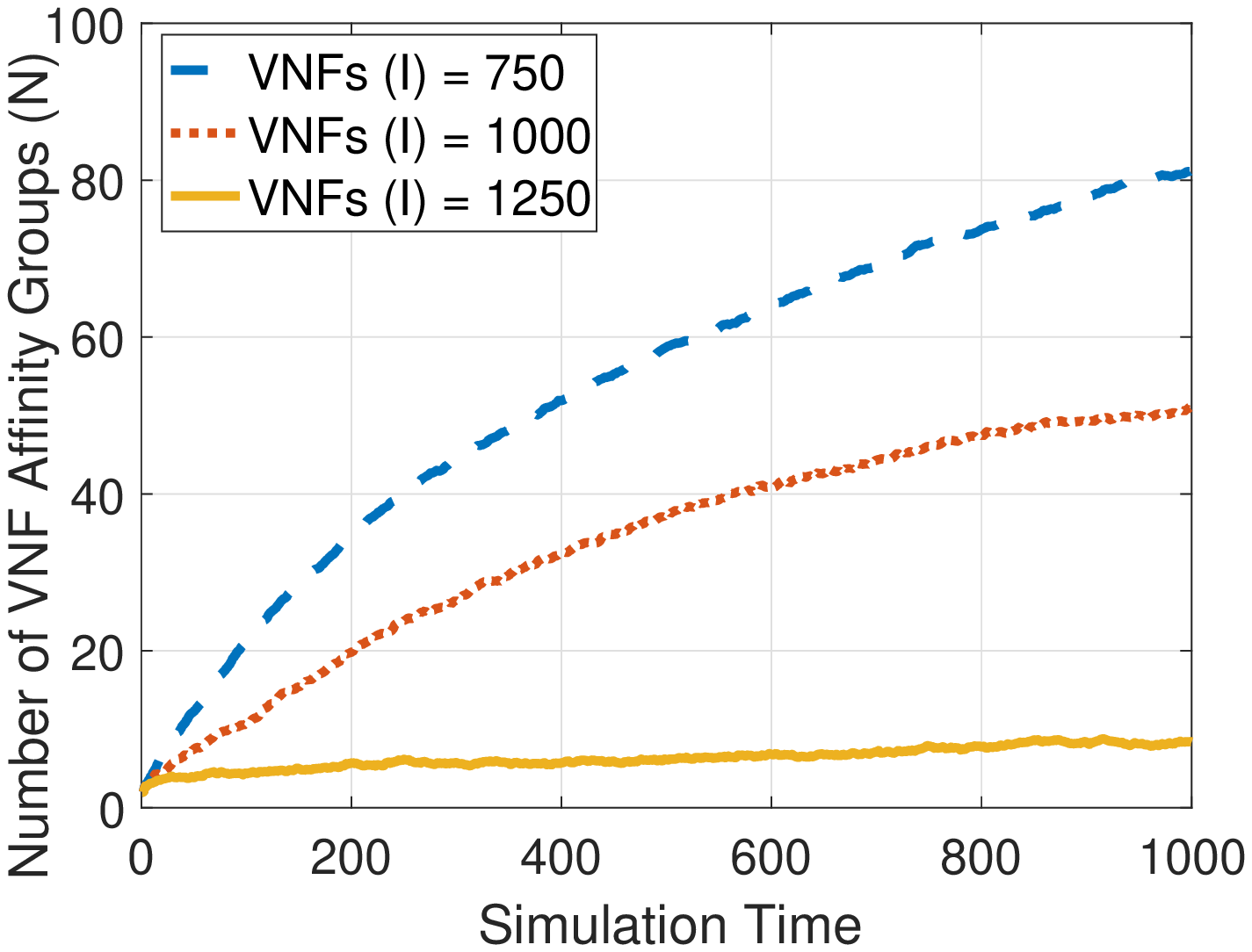}
 }
 \subfigure[Monitoring Frequency ($\delta$) varying $I$]
 {
	\label{fig:multiDelta_I}        
    \centering
	\includegraphics[width=0.3\textwidth]{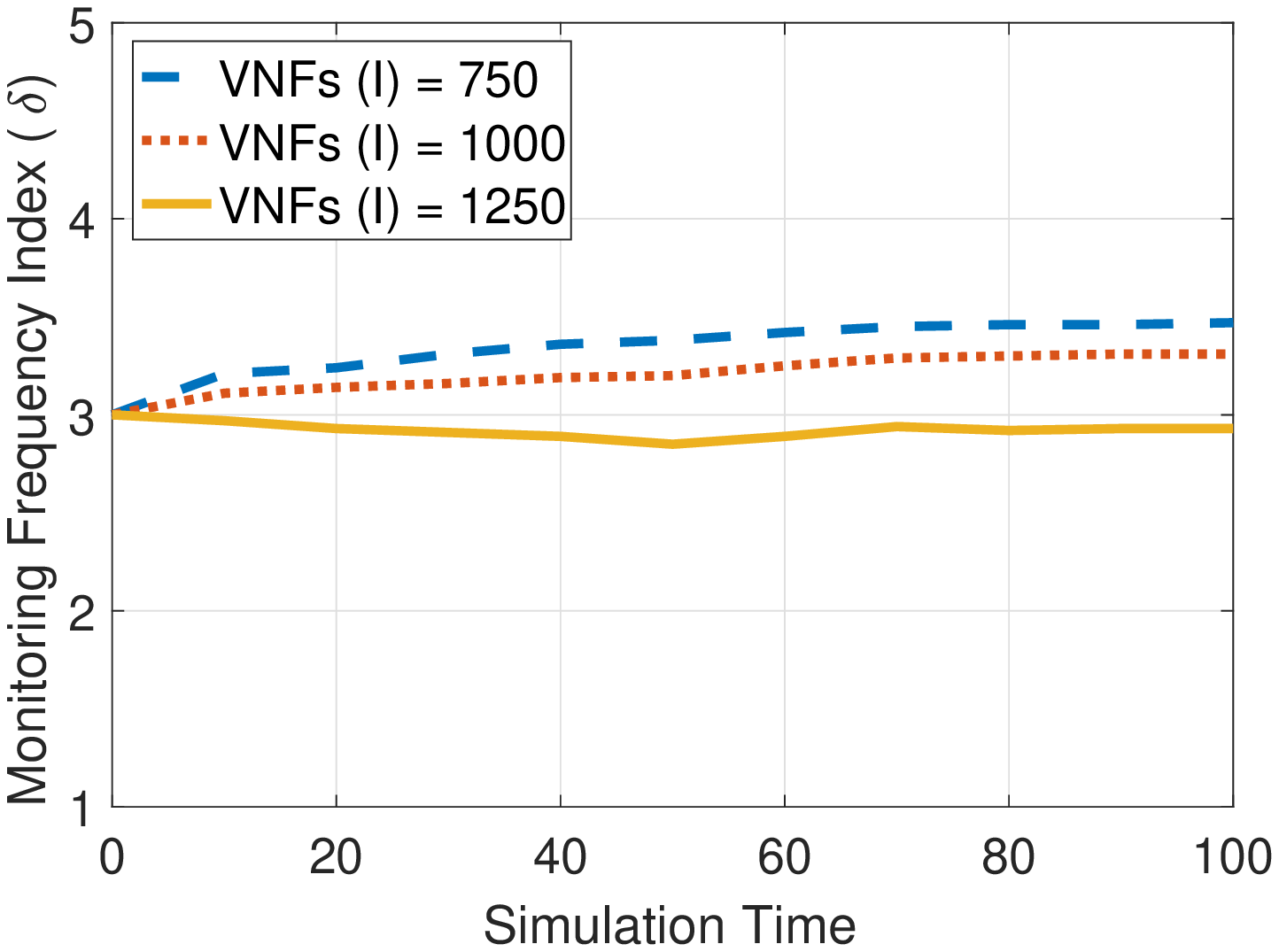}
 }
 \subfigure[Surveillance Epoch length ($\omega$) varying $I$]
 {
	\label{fig:multiOmega_I}        
    \centering
	\includegraphics[width=0.3\textwidth]{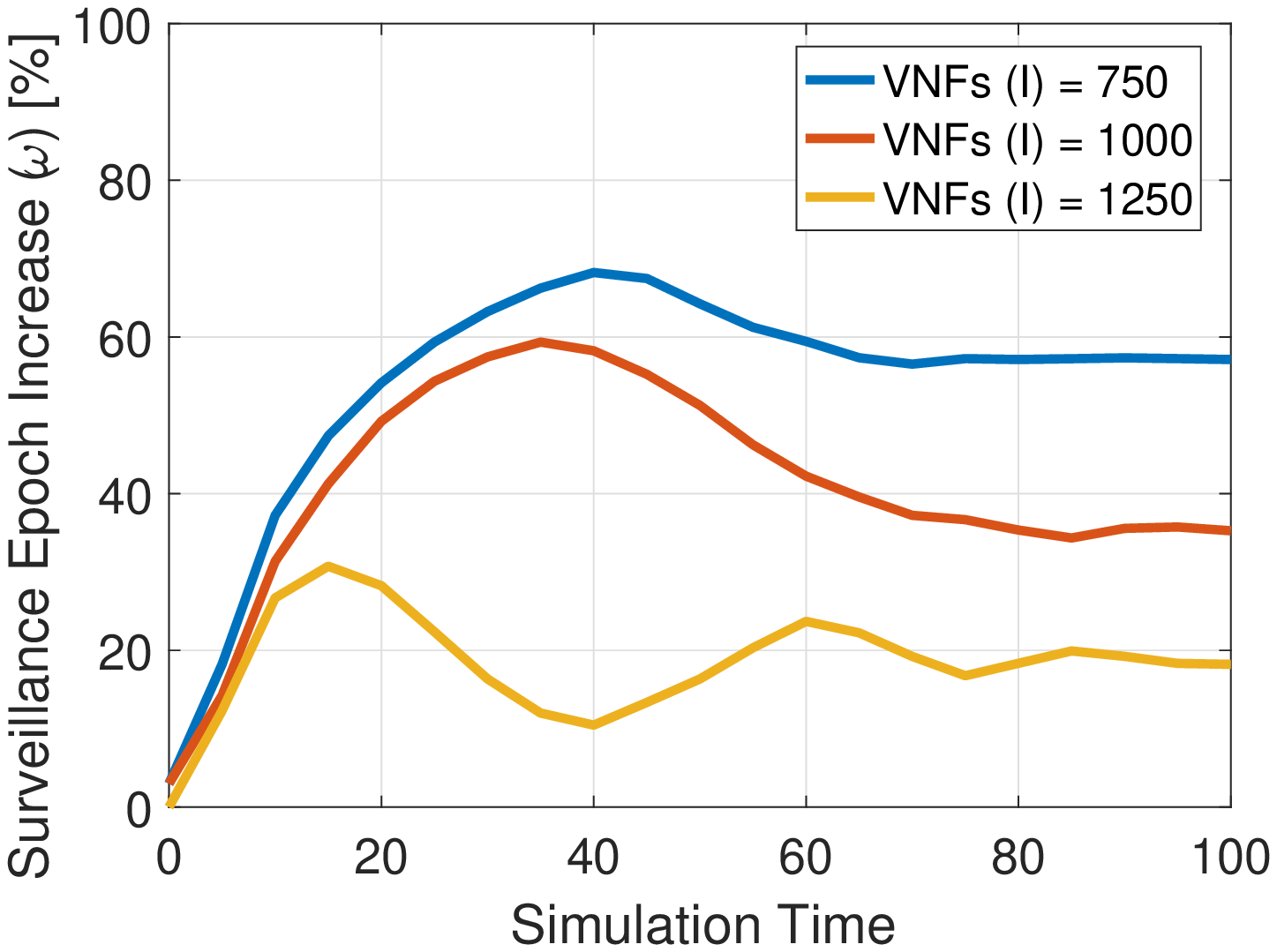}
 }
 \subfigure[VNF Affinity Groups (N) increasing $\sigma$]
 {
	\label{fig:multiN_var}        
    \centering
	\includegraphics[width=0.3\textwidth]{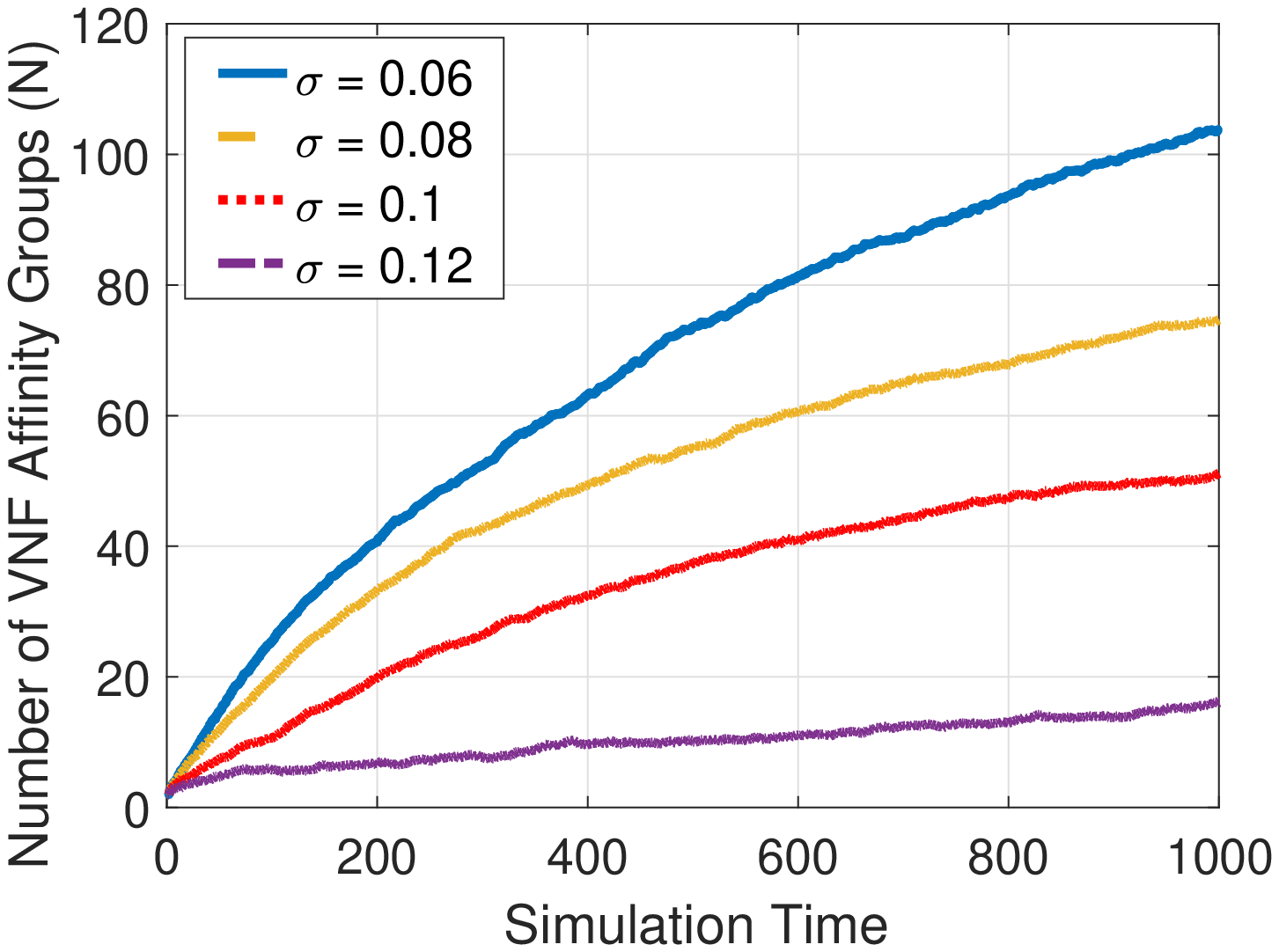}
 }
 \subfigure[Monitoring Frequency ($\delta$) increasing $\sigma$]
 {
	\label{fig:multiDelta_var}        
    \centering
	\includegraphics[width=0.3\textwidth]{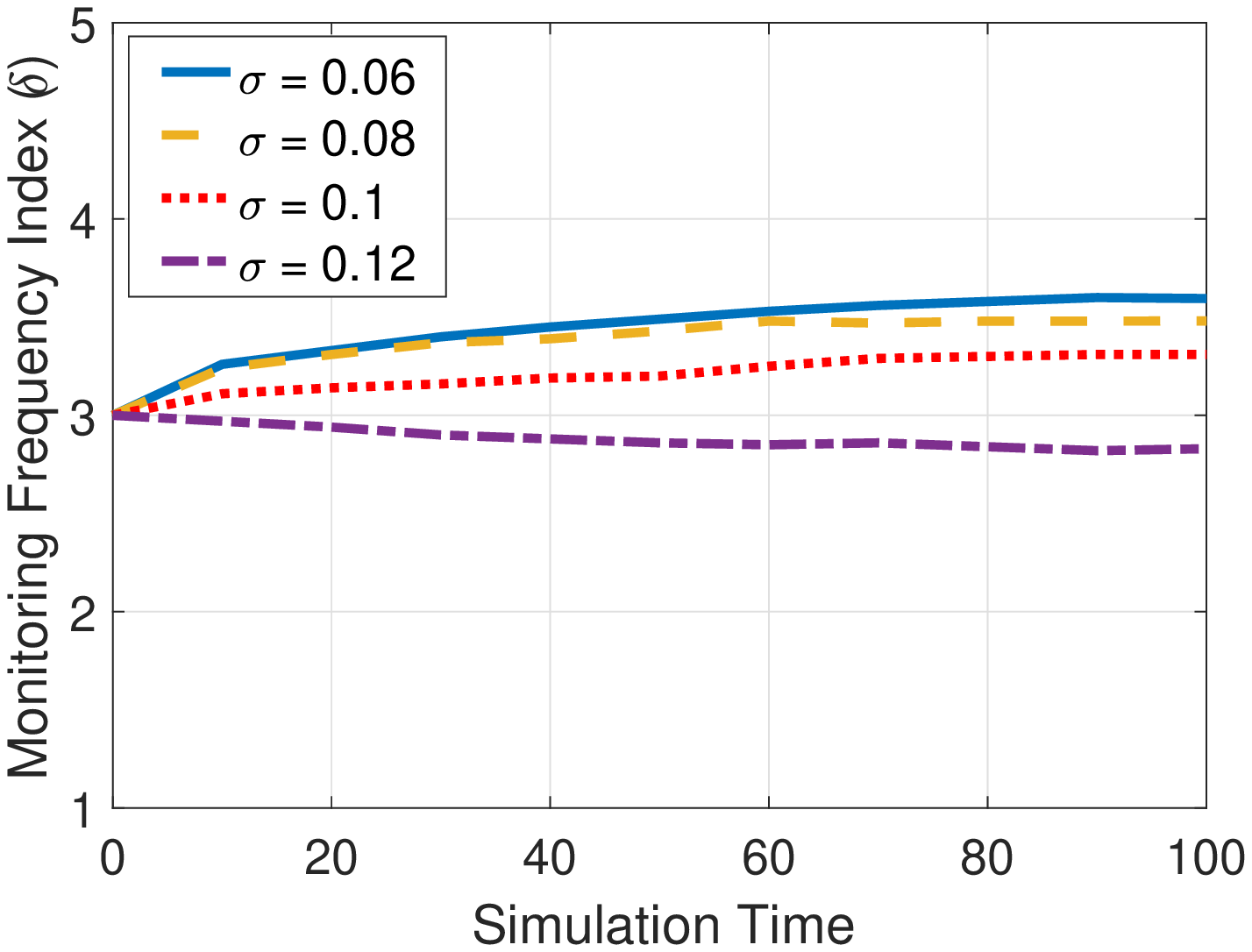}
 }
 \subfigure[Surveillance Epoch length ($\omega$) increasing $\sigma$]
 {
	\label{fig:multiOmega_var}        
    \centering
	\includegraphics[width=0.3\textwidth]{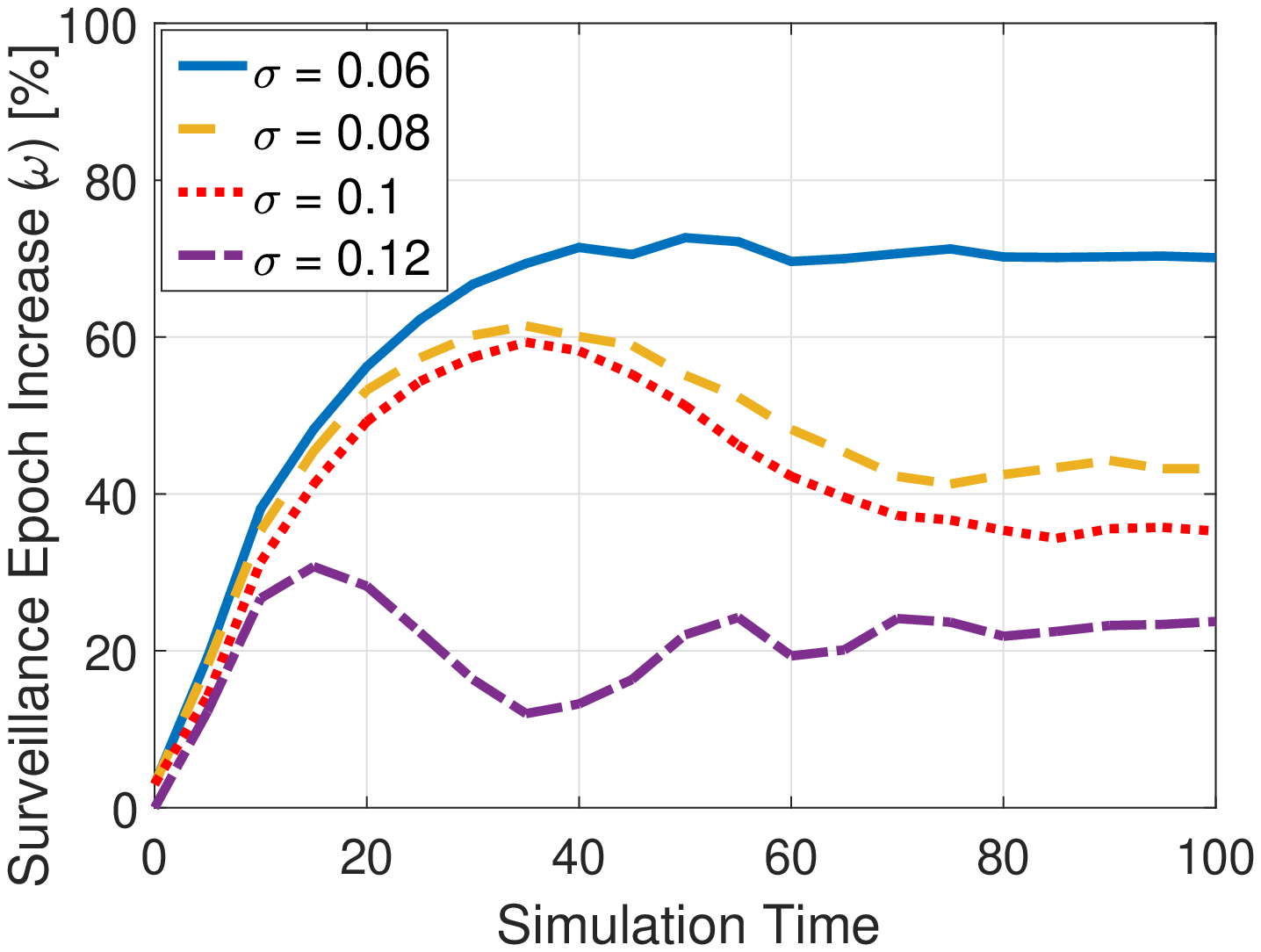}
 }
 \caption{Evaluation of adaptation parameters.}
 \label{fig:fig1}
 \end{figure*}

\vspace{-5mm}
\subsection{Simulation setup}
The NFV system service orchestration is performed for a huge number of VNF instances. Given the unavailability of such a complex real testbed, \change{we assess the performance of our approach by means of synthetic simulations} taking into account as baseline the real values offered by OpenEPC VNFs and, at the same time, shedding the light on the impact of a large number of VNFs deployed on different compute nodes.

VNFs profiles are built based on a Pareto random distribution using the values listed in Table~\ref{tab:kpis_vnf}. The long-tail effect of the random distribution is handled with a cap to limit VNF resource utilization to $100\%$. \change{Once VNF baseline profiles are defined, at every time slot $t$ a VNF instance is executed and VNF profile KPIs are generated and collected based on a normal distribution with the VNF baseline profile as mean, and variance $\sigma$ based on the considered scenario. If not differently stated, used simulation parameters are listed in Table~\ref{tab:system_param}.}

\begin{table}[h!]
\caption{Simulation parameters}
\label{tab:system_param}
\vspace{-2mm}
\scriptsize
\centering
\begin{tabular}{|c|c||c|c|}
\hline
\textbf{Parameters} & \textbf{Values} & \textbf{Parameters} & \textbf{Values}\\
\hline
VNF Profiles ($I$) & \!\!\!$1000$\!\!\! & VNF Baseline Profiles & $5$\\
VNF Profile KPIs ($Z$)\!\!\! & $3$ & VNF Profile variance ($\sigma$) & $0.1$\\
Surveillance Epoch ($\omega$)\!\!\! & \!\!\!$500$t\!\!\!& Monitoring Interval ($1/\delta$) &\!\!\! [$2,5,10,20,50$]\!\!\!\\
Q-learning ($\beta$) & $0.5$ & Simulation time & $10^7$t\\
Q-learning ($\psi$) & $0.9$ & Q-learning ($\phi$) & $0.5$\\
\hline
\end{tabular}
\end{table}

\subsection{System parameters evolution}
\change{We study and discuss the evolution of the system parameters as well as their consistent effects on the overall system efficiency from two different perspectives: $i$) the VNF placement and Quality-of-Decisions and $ii$) the VNF monitoring load.}
 
The main finding of our simulation campaign lies on the concept of VNF profile variability. VNF profile variability plays a key-role in the VNF placement and then in the overall Quality-of-Decisions of the CMS. VNF profiles exhibiting significant profile deviations may result in relevant performance degradations, as the system must detect unexpected behaviours and promptly react. We study the evolution through three different adaptive parameters: the number of VNF affinity groups $N$, the monitoring frequency $\delta$ and the surveillance epoch length $\omega$, as shown in Fig.~\ref{fig:fig1}.

The number of affinity groups $N$ could unveil interesting aspects. \change{z-TORCH automatically tailors the affinity group characteristics onto specific VNF profile properties, given that no VNF profile deviations occur. In other words, as soon as the unsupervised binding affinity process successfully identifies the VNF affinity groups (keeping low the risk of profiling failure), the granularity of such a process will be reduced (i.e., more groups will be defined) in the next decisional slot to increase the accuracy of the binding. Conversely, when a failure in the binding process is detected (due to unexpected changes), the granularity of the VNF affinity groups is automatically enlarged leading to a fewer number of groups (with larger scopes). This clear evidence is provided by Figs.~\ref{fig:multiI_N} and~\ref{fig:multiN_var}. 
When the VNF profile variance $\sigma$ is low or when a few VNF profiles are considered, the accuracy of the unsupervised binding affinity process is large enough to allow our solution to increase (quickly) the number of considered affinity groups. This leads to a more efficient calculation and low probability of failure when placing VNFs based on their profile (i.e., assigned affinity group). However, when the variability becomes consistent (or the number of VNF profiles grows), the binding failures (due to VNF profile deviation) might affect the accuracy of the process that automatically enlarges the scope of each single affinity groups so as to account for unexpected variability while reducing their total amount.
}

\begin{figure*}[t]
 \centering
 \subfigure[VNF Affinity binding using e$k$m]
 {
	\label{fig:ekm}        
    \centering
	\includegraphics[width=0.3\textwidth]{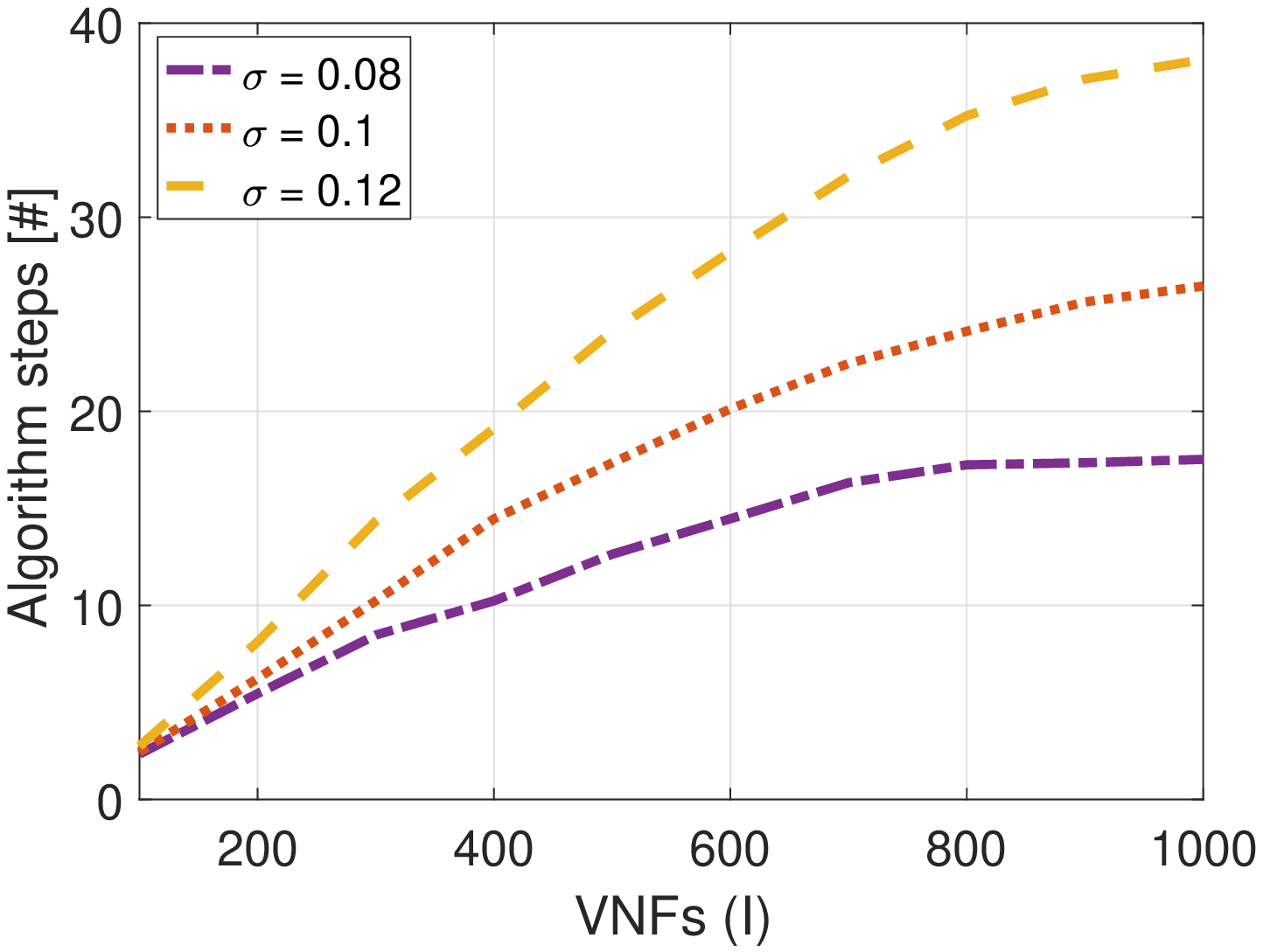}
 }
 \subfigure[Proactive VNF Placement]
 {
	\label{fig:proact}        
    \centering
	\includegraphics[width=0.3\textwidth]{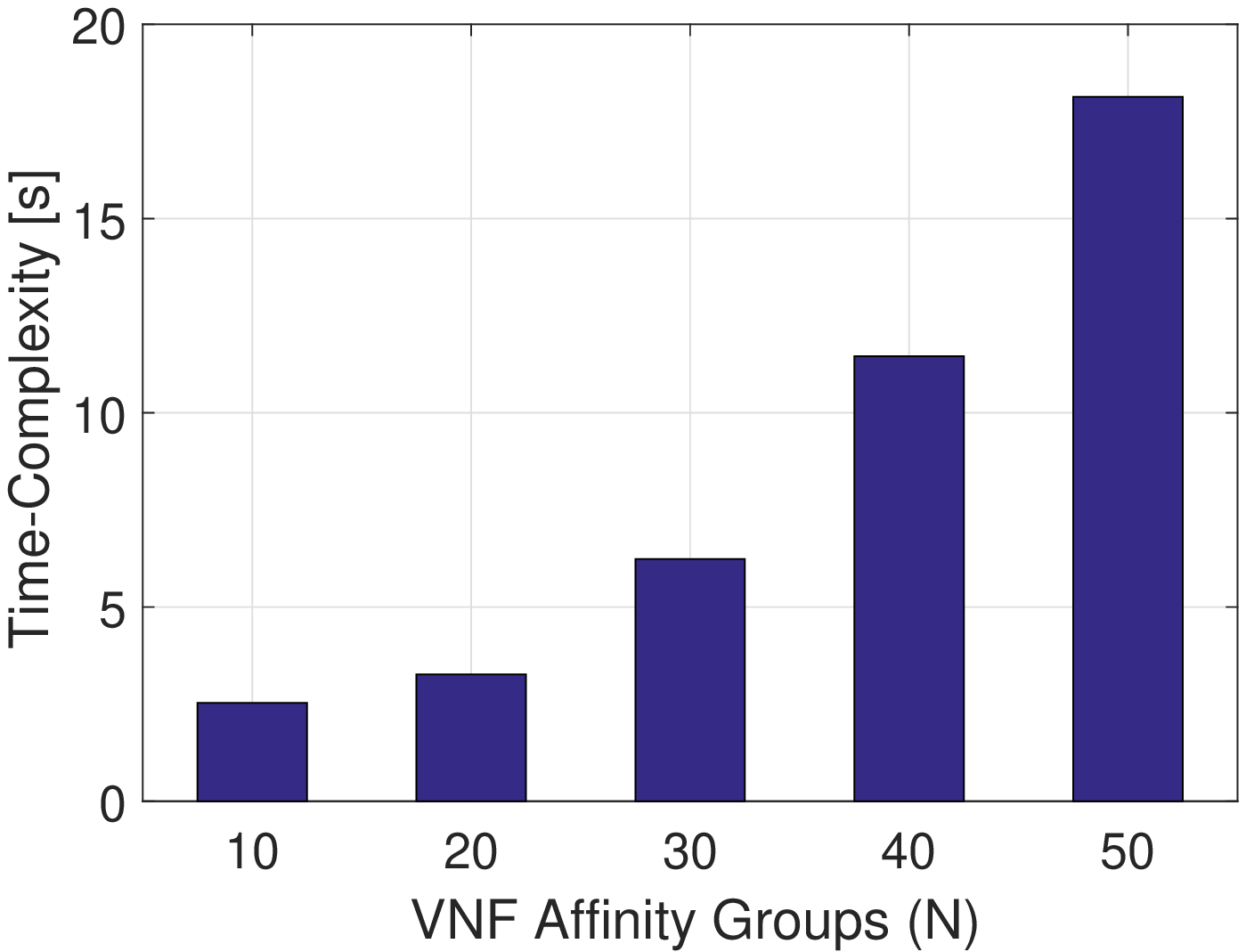}
 }
 \subfigure[VNF placement using AaVS]
 {
	\label{fig:aavs}        
    \centering
	\includegraphics[width=0.3\textwidth]{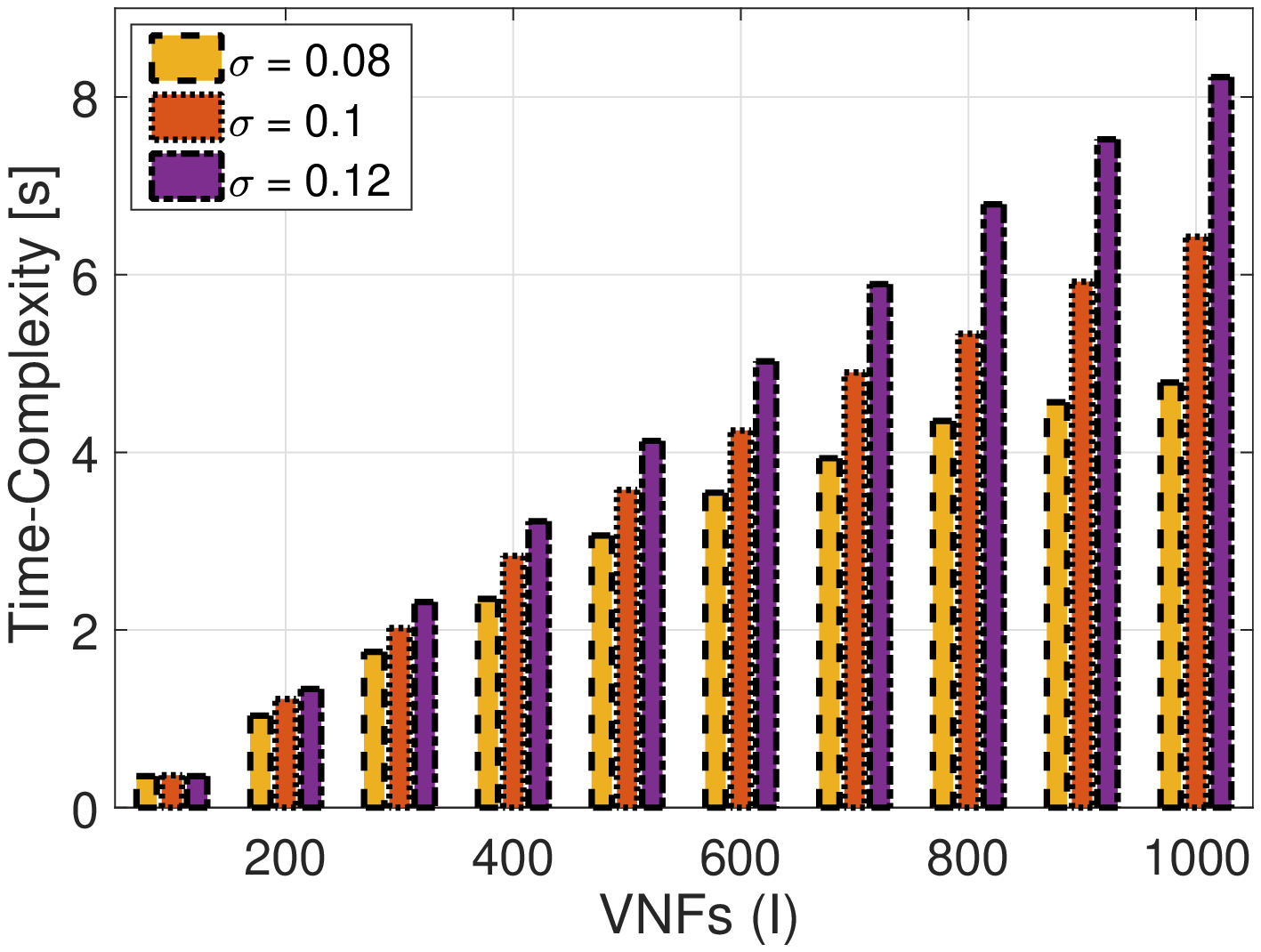}
 }
 \caption{Solution complexity analysis.}
 \label{fig:fig2}
 \end{figure*}
\change{Another important feature of z-TORCH is the monitoring load which is directly triggered by the monitoring frequency $\delta$. In our simulations, we consider a fixed set of $5$ frequency intervals, where the largest index ($5$) results in a very low monitoring load. Figs.~\ref{fig:multiDelta_I} and ~\ref{fig:multiDelta_var} show the evolution of the monitoring index. When the statistical variance $\sigma$ or the number of VNF instances is low, the system reduces the monitoring burden, on average. This is due to a more stable system state and a limited risk of profiling failure. On the other side, when the number of VNF profile deviations grows, the monitoring load needs to be promptly adapted incurring in more monitoring messages.}

The last parameter is the surveillance epoch length $\omega$, which has a two-fold aspect: $i$) it might significantly boil down the complexity of our solution by delaying the next decisional time for making LCM decisions and $ii$) it impacts on the number of monitoring information accounted for the next binding affinity group operation. This parameter is driven by the Q-learning approach, as explained in Section~\ref{sect:qlearning}.
In Figs~\ref{fig:multiOmega_I} and~\ref{fig:multiOmega_var}, we show the effect of the VNF profile variability $\sigma$ and the number of considered VNF profile instances $I$. 
\change{High VNF profile variance and huge number of VNF profile instances result in more unstable behaviours requiring short monitoring surveillance epochs and, in turn, more decisional times. When the variability of the VNF instances is limited, the surveillance epoch length on average increases (and in turn reduces the complexity of the decisional mechanisms) and stabilizes.}

\subsection{Complexity and time performance}
\change{While the adaptiveness of z-TORCH allows to promptly react to unexpected changes in the VNF profiles and to reduce the monitoring load, here we show the cost in terms of complexity of our novel mechanism for each novel algorithm.
}
In Fig.~\ref{fig:ekm}, we show the number of steps of enhanced $k$-means (e$k$m) algorithm needed to converge. Notably, the variability of the VNF profiles might affect the complexity of the algorithm. However, the curves exhibit a sub-linear dependency on the number of VNF profile instances, which makes our algorithm suitable even for crowded VNF environment.

\change{We next analyze the time complexity in terms of seconds for the Proactive VNF Placement problem solution using a commercial solver, namely IBM ILOG CPLEX. Specifically, we run our algorithm on a dual Intel(R) Xeon CPU $2.40$GHz $4$-cores and $16$GB RAM. Fig.~\ref{fig:proact} shows the time complexity in terms of elapsed seconds when considering a different number of VNF affinity groups. As expected the complexity of such solution grows exponentially with respect to the number of affinity groups (centres of gravity) due to the NP-Hardness property of the optimization problem described in Section~\ref{sect:vnf_placement}. However, in realistic environments the number of VNF affinity groups is low when compared to the number of VNF profile instances, making our approach valid and reasonable.}

Last, we show the time complexity performance of the VNF placement algorithm, namely AaVS, as described in Section~\ref{sect:vnf_placement}. In Fig.~\ref{fig:aavs} we depict complexity results when applied different VNF profile variance $\sigma$ and VNF profile instances $I$. \change{Interestingly, high values of $\sigma$ exacerbates the growing rate of the complexity but still showing a sublinear behavior, which in our test never exceeds $9$ seconds. We can conclude that AaVS is easily applied for realistic scenarios where the number of VNF instances may dramatically grow.} 

\subsection{z-TORCH: advantages and limitations}
 
Due to the lack of existing solutions addressing jointly both optimal placement (Quality-of-Decisions) and monitoring load minimization, we compare the performance of z-TORCH against a legacy approach, wherein optimal VNF placement decisions are taken every decisional time without exploiting machine-learning solutions. We call this benchmark as \emph{Instant Placement}. Additionally, to evaluate the goodness of our solution, we develop an optimal VNF placement solution, namely \emph{Optimum}. This solution possesses a God-knowledge of the future VNF profile deviations. Therefore, it can calculate the optimal VNF placement (for each decisional time) in order to minimize the overall VNF migrations in the future. We denote the performance difference between our approach and the optimal one as \emph{Regret}, following the online decisional algorithms terminology.

\begin{figure}[t]
 \centering
 \subfigure[Quality-of-Decisions]
 {
	\label{fig:qod}        
    \centering
	\includegraphics[width=0.35\textwidth]{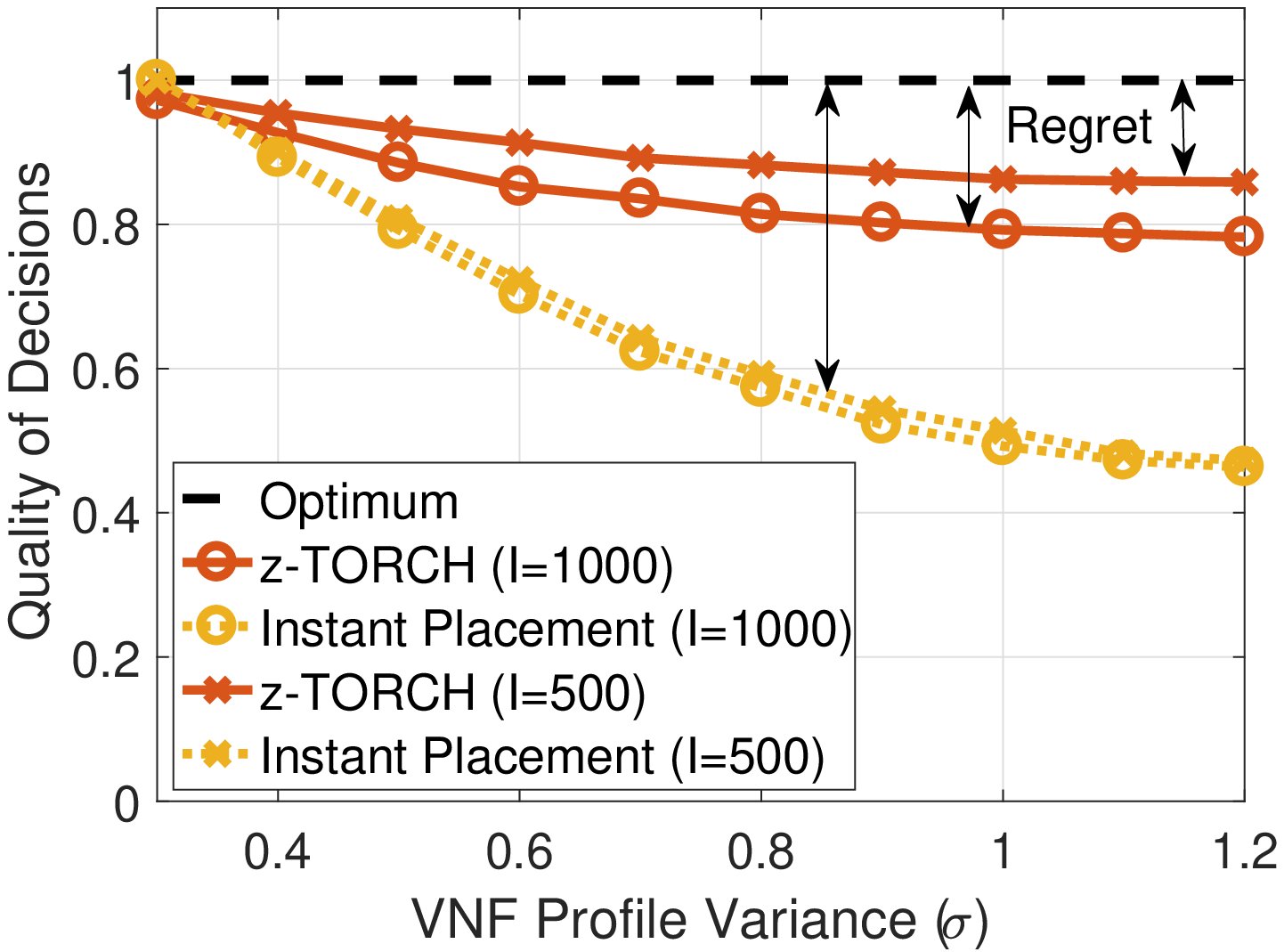}
 }
 \subfigure[Monitoring Load]
 {
	\label{fig:mon_load}        
    \centering
	\includegraphics[width=0.35\textwidth]{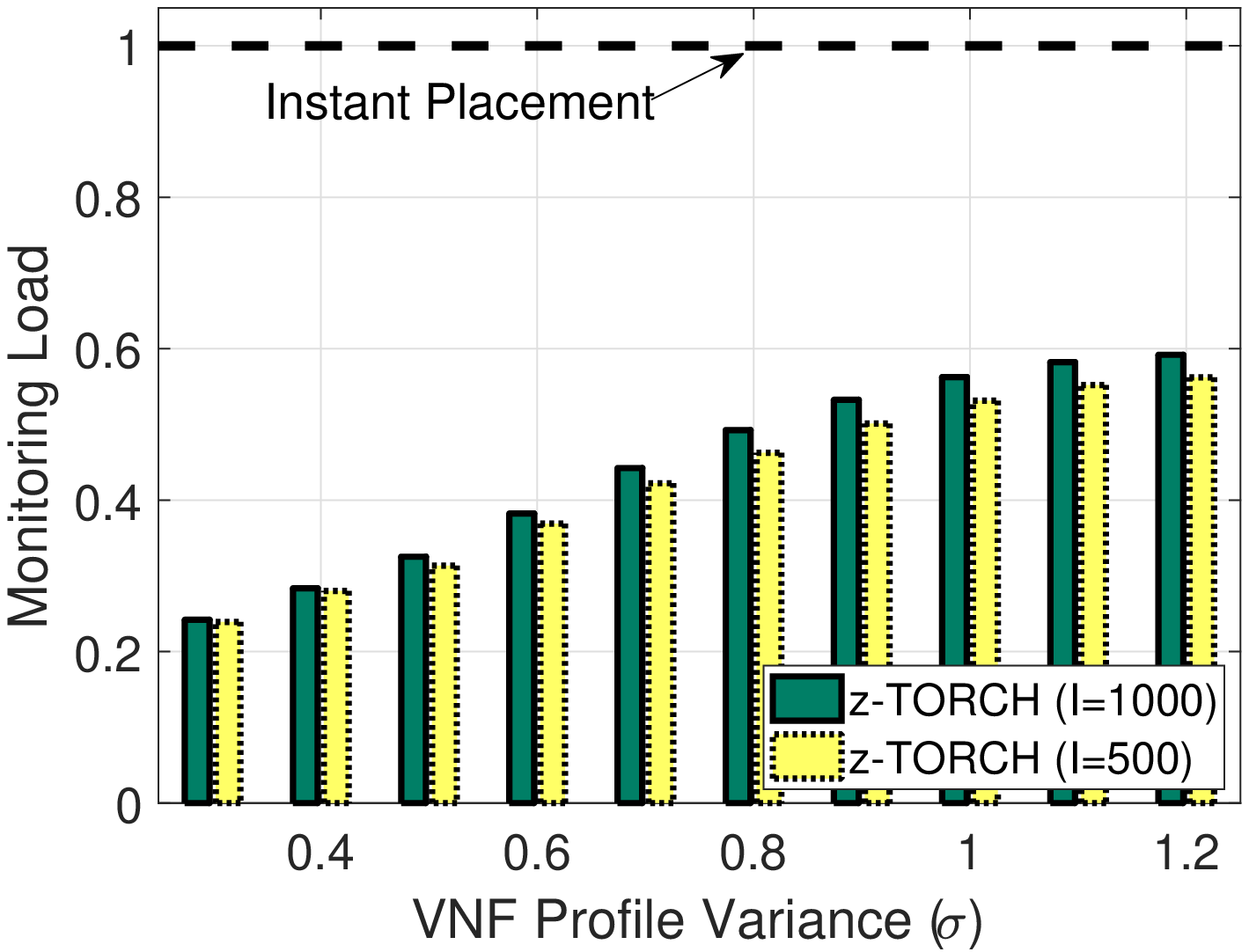}
 }
 \caption{z-TORCH: Placement and monitoring performance.}
 \label{fig:fig3}
 \end{figure}

We evaluate our approach in terms of Quality-of-Decisions (QoD) assuming that the \emph{Optimum} policy takes the best decision, i.e., QoD $=1$. We use then the number of migrations performed by the optimal policy as benchmark, and we calculate the number of VNF migrations exceeding the benchmark. Resulting QoD is the ratio between the optimal number of migrations and the number of migrations required by each solution. In Fig.~\ref{fig:qod}, we show the QoD results while varying the VNF profile variance $\sigma$ for two different scenarios with $500$ and $1000$ VNF profile instances. Interestingly, when the variance is very low, i.e., VNF profiles are predictable and stable, the Instant Placement solution slightly outperforms z-TORCH. This is due to the initial training phase in which z-TORCH needs to adapt and stabilize. When the VNF profile variance increases, z-TORCH shows a near-optimal results (up to $88.6\%$) almost doubling the performance of the Instant Placement solution.

Last, we show the monitoring load analysis when z-TORCH is in place. In this case we only compare against the Instant Placement, as the optimum solution is executed only once. Instant Placement can be considered as the worst case since it needs monitoring information every sample point ($\delta$). Therefore, we normalize the number of monitoring messages needed by z-TORCH by the ones needed by the Instant Placement solution. Results are depicted in Fig.~\ref{fig:mon_load}. The larger the variability of VNF profiles increases, the higher the monitoring load. This is due to a number of VNF profile deviations, which must be controlled through more monitoring information. However, the monitoring load seems to stabilize around $50-60\%$ even for significant variance values $\sigma$. 

This confirms that z-TORCH outperforms legacy solutions while showing near-optimal performance at low monitoring costs. Nonetheless, considered solutions (Instant Placement and Optimum) requires a huge complexity making them not suitable for being executed in an affordable time.

\section{Deployment Considerations}
\label{sec:deployment}
In this section, we will provide insights at various deployment and implementation considerations  \textcolor{blue}{of our proposed method} with respect to standard ETSI NFV MANO system \cite{etsiMano}  \textcolor{blue}{and open source MANO projects like Open Network Automation Platform (ONAP) \cite{onap} and Open Source MANO (OSM) \cite{osm}}.

The ETSI NFV MANO system,, which is a standard CMS for NFV based environment, is composed of three main functional blocks namely the Virtualized Infrastructure Manager (VIM), VNF Manager (VNFM) and NFV Orchestrator (NFVO). The ETSI NFV MANO system is designed to manage and orchestrate virtualized resources in an NFV Infrastructure (NFVI) such as virtualized compute, network, storage, memory etc via the VIM. It also manages the individual VNFs that are deployed over the NFVI via the VNFM. The NFVO is designed to perform resource orchestration and service orchestration, where the service meant here is the Network Service (NS) that is formed by the concatenation of multiple relevant VNFs to provide a composite network service. In other words the VIM, VNFM and NFVO constitute the CMS. There are no specific proposals as to how the monitoring system will be integrated in the NFV MANO system. It is implied that the VIM, VNFM and the NFVO will monitor their respective layers for performance and fault management and take relevant LCM decisions as per the logic local to the respective functional block. There is also a requirement to monitor the MANO functional blocks for its own performance/fault management and that there is indeed a requirement to have a monitoring entity i.e., MANO Monitor, with which all the three MANO functional elements will interact with \cite{etsiIfa021}. However, there is no specific architectural proposal. In view of the prevailing understanding, there are thus two layers of monitoring for performance/fault management; Layer\-1 is for the monitoring of the virtualized infrastructure and resources, while Layer\-2 is for the monitoring of the MANO functional blocks themselves. 
In this regard we propose two possible deployment options for integrating a monitoring system within the ETSI NFV MANO framework that can then be leveraged by the proposed z-TORCH method.

\begin{figure}[t!]
 \includegraphics[clip,width=\linewidth]{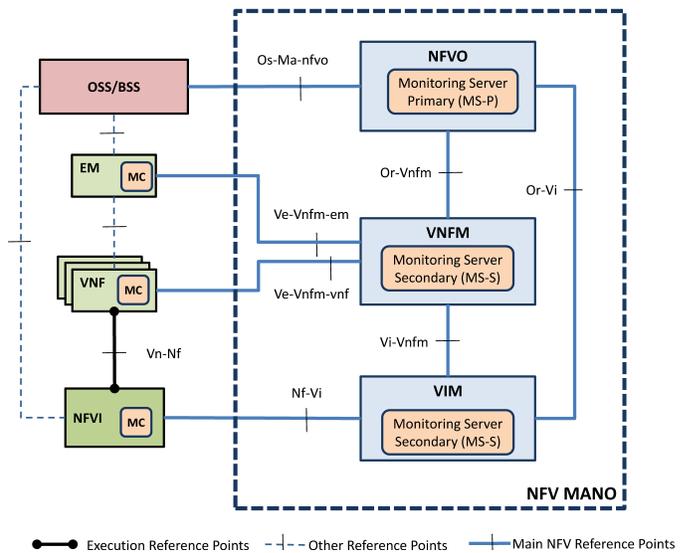}
 \caption{NFV MANO system with integrated monitoring system (Distributed).}
 \label{fig:embod1}
 \end{figure}

\subsection{Deployment Option 1}

This option is illustrated in Fig.~\ref{fig:embod1}, where the MS is integrated within each MANO functional blocks while the MCs are deployed within the virtualized infrastructure/resources. As explained above, the MC will be configurable by the MS. The key difference is that due to the distribution of the MS inside the MANO functional blocks, the MS within the NFVO is the Primary MS (MS\-P), while the MS inside the VIM and VNFM are the Secondary MS (MS\-S). The MS\-Ss can independently monitor, collect, analyze data from the functional block respective layer. For example, the MS\-S within the VNFM will be able to deploy and configure MC instances inside the VNF instance(s) and will also independently collect and locally analyze monitored data from these MCs. Based on the analysis of the monitored data, the VNFM can take VNF specific LCM decisions as per the policy/decision logic local to the VNFM. Similarly the MS-S inside the VIM will deploy and configure MC within the virtualized/non-virtualized infrastructure resources (e.g., compute, network, memory, storage) and monitor and manage them as per its local policy/decision logic. 
However, the LCM decisions taken by the VIM and/or VNFM must be validated by the NFVO as the latter has an overview of the overall NS that is composed of several VNFs managed by possibly different VNFMs and deployed over possibly different VIM platforms. 

 Owing to the level and centrality of the NFVO in the LCM decision process; the MS-P is integrated within NFVO. The MS-P does not deploy/configure/monitor any specific MCs but it monitors and configures the MS-S instances in VNFM and VIM. The MS-P may override any configuration parameter within the MS-S instances at any time. Our proposed method shall typically run inside MS\-P and based on the feedback it receives from MS-S will (re)compute and (re)adjust the values of $\omega$ and/or $\delta$ and/or $t$ for the specific MS\-S instances. Based on these values, the MS\-S will (re)configure the MC instances within their respective monitoring domain. The MS\-P will also configure the MS-S with the KPIs to monitor and can change the configuration parameters of the MS-S any time. The MS\-P, based on the inputs received from the MS\-S will forward them to the analysis engine (AE). The AE after analyzing the data send the results to the decision engine which will take appropriate decision on LCM, recompute the necessary configuration parameters for the MS\-S instances and push them over the respective standard reference points i.e., Or\-Vi and Or\-VNFM reference points. Please note that the AE and DE components and their inter-relationship with themselves and the MS\-P is similar to what is shown in Fig.~\ref{fig:cms}.Our proposed method can either run in the MS-P or the AE and the AE then provide the recommended configurations parameters to the MS-S. 

\begin{figure}[t!]
 \includegraphics[clip,width=\linewidth]{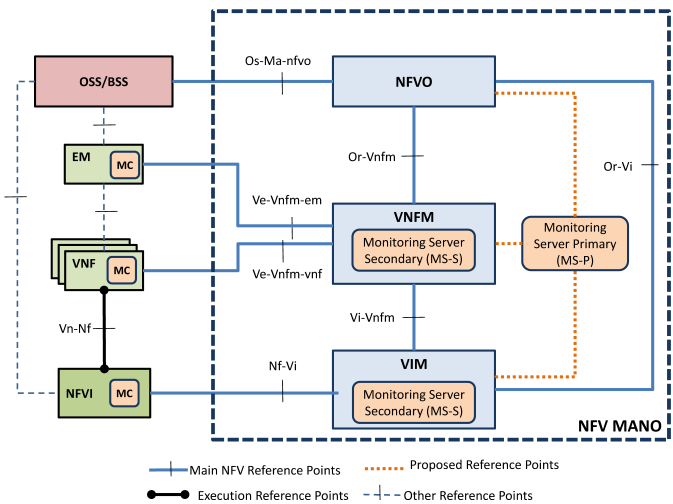}
 \caption{NFV MANO system with integrated monitoring system (Centralized).}
 \label{fig:embod2}
 \end{figure}

\subsection{Deployment Option 2}

This option is depicted in Fig.~\ref{fig:embod2}. In this deployment, the MS-P is a central entity that interacts with the MS-S located in VNFM and VIM functional blocks. The NFVO does not carry a MS-S as it will interact with the external MS-P. The method described here is implemented in MS-P that will then be used to compute the relevant configuration parameters for the MS-C, which in turn will configure the MCs of their respective domains. In this case the NFVO, which carries the AE and the DE (see Fig.~\ref{fig:cms}) will inform the MS-P of its LCM decision and also the identities of the VNFs and NSs that has been affected, and based on this information the MS-P will (re)calculate the relevant configuration parameters and push then to the MS-Ss so that they can configure the MCs within their respective layer. It is also possible that the MS-P may derive separate configuration values for the MS-S. This will make the MC at the NFVI and VNF level to use different monitoring configuration. 

 \textcolor{blue}{In addition to the above two proposed deployment options, it is worth mentioning that there are open source MANO projects like ONAP and OSM that are in various stages of development. Having a credible monitoring system for data collection is integral to the design of these frameworks. For example, OSM has a Monitoring Module (MON) which interfaces with 3rd party monitoring systems, and is used for pushing monitoring configuration updates to external monitoring systems while steering a limited set of actionable evens into the Service Orchestrator \cite{osm}. ONAP on the other hand has a more elaborate design for this purpose. In ONAP framework, there is a dedicated DCAE platform that consists of several functional components like, Collection Framework, Data Movement, Storage Lakes, Analytic Framework, and Analytic Applications \cite{dcae}. The Collection Framework within the DCAE enables the collection of various types of data such as, event data for monitoring the health of the managed environment, data to compute the key performance and capacity indicators necessary for elastic management of the resources, and granular data needed for detecting network and service conditions\cite{dcae}. The collected data is then processed by the Analytic Framework for anomaly detection, capacity monitoring, congestion monitoring, or alarm correlation etc. The Analytics Framework also enables agile development of analytic applications, and from this perspective is more suitable for the implementation of z-TORCH method. This is the next step, where we are evaluating the features and capabilities of the DCAE platform for testing and evaluating z-TORCH in a real test environment.}

\section{Conclusions}
\label{sec:conclusion}
In this work, we have designed an automated solution, namely z-TORCH, performing joint NFV orchestration and monitoring re-configuration operations without requiring human intervention. We have built our solution based on machine-learning approaches. In particular, we have proposed an unsupervised binding affinity solution to study and profile VNF KPIs. This has allowed us to proactively place VNFs into compute nodes pursuing the Quality-of-Decisions (QoD) maximization and, in turn, the decisional complexity minimization. In addition, z-TORCH automatically adapts the VNF monitoring load according to VNF profile time variations. 

The main characteristics of our proposed z-TORCH solutions can be summarized as follows: $i$) an unsupervised system in charge of profiling VNF KPIs based on previous monitoring information, $ii$) a proactive VNF placement based on pre-calculated affinity groups, $iii$) an adaptive monitoring load control to minimize the overhead of monitoring information. NP-Hardness proofs and heuristics algorithms are introduced to make our framework practical and implementable. An exhaustive simulation campaign is carried out to validate our solution against a legacy system showing that z-TORCH can achieve near-optimal results at very limited monitoring costs.

\change{As a next step, we will evaluate the features and capabilities of open source MANO projects like ONAP for testing and evaluating z-TORCH in a real test environment.}\vspace{-2mm}

\section{Acknowledgments}
This work has received funding from the European Unions Horizon 2020 research and innovation programme under grant agreement No 761536 (5G-Transformer project).

\bibliographystyle{IEEEtran}
\bibliography{IEEEabrv,references}

\end{document}